\documentclass[11pt,a4paper]{elsarticle}

\usepackage{graphicx}
\usepackage{graphics}
\usepackage{amsmath}
\usepackage{times}
\usepackage{psfrag}
\usepackage{amssymb}
\usepackage{subfigure}
\usepackage{color}
\usepackage[dvipsnames]{xcolor}
\usepackage{bm}

\newcommand{\tpf}[1]{{\color{#1}\mbox{%
			\protect\rule[0.5ex]{1.5mm}{0.25mm}\hspace{0.75mm}\protect\rule[0.5ex]{1.5mm}{0.25mm}\hspace{0.75mm}\protect\rule[0.5ex]{1.5mm}{0.25mm}}}}

\newcommand{\tppp}[1]{{\color{#1}\mbox{%
			\hspace{0.75mm}\protect\rule[0.5ex]{0.5mm}{0.25mm}\hspace{0.5mm}\protect\rule[0.5ex]{0.5mm}{0.25mm}\hspace{0.5mm}\protect\rule[0.5ex]{0.5mm}{0.25mm}\hspace{0.5mm}\protect\rule[0.5ex]{0.5mm}{0.25mm}\hspace{0.5mm}\protect\rule[0.5ex]{0.5mm}{0.25mm}\hspace{0.5mm}\protect\rule[0.5ex]{0.5mm}{0.25mm}
}}}
\begin{document}
	
	\begin{frontmatter}
		
		\title{Identification of Secondary Resonances of Nonlinear Systems using Phase-Locked Loop Testing}

		\author{Tong~Zhou\corref{cor1}}
		\ead{tong.zhou@uliege.be}
  
		\author{Ga{\"e}tan~Kerschen}

		\address{Space Structures and Systems Laboratory, Aerospace and Mechanical Engineering Department, University of Liege, Belgium}  
  
        \cortext[cor1]{Corresponding author}

		\begin{abstract}
  
One unique feature of nonlinear dynamical systems is the existence of superharmonic and subharmonic resonances in addition to primary resonances. In this study, an effective vibration testing methodology is introduced for the experimental identification of these secondary resonances. The proposed method relies on phase-locked loop control combined with adaptive filters for online Fourier decomposition. To this end, the concept of a resonant phase lag is exploited to define the target phase lag to be followed during the experimental continuation process. The method is demonstrated using two systems featuring cubic nonlinearities, namely a numerical Duffing oscillator and a physical experiment comprising a clamped-clamped thin beam. The obtained results highlight that the control scheme can accurately characterize secondary resonances as well as track their backbone curves.
A particularly salient feature of the developed algorithm is that, starting from the rest position, it facilitates an automatic and smooth dynamic state transfer toward one point of a subharmonic isolated branch, hence, inducing branch switching.
  
		\end{abstract}

		
	\end{frontmatter}

%==================================================================
\section{Introduction}
%==================================================================

%\subsection{Background: the need for nonlinear analysis}

Experimental modal analysis is the most common approach for identifying the dynamical characteristics of a mechanical system. 
Specifically, the objective is to determine the modal parameters, namely the vibration modes, resonance frequencies and damping ratios, from the analysis of free or forced vibrations. At the root of experimental modal analysis is the assumption of linear dynamical behavior. However, nonlinear phenomena are more and more frequently encountered during vibration testing of modern engineering structures \cite{kerschen2006}. 
Typical examples include commercial aircraft \cite{goge2007}, satellites \cite{noel2014}, turbomachinery \cite{schwarz2020} and microelectromechanical systems (MEMS) \cite{frangi}. This is why there is currently a burgeoning interest in developing methodologies that can tackle nonlinear behaviors.

%\subsection{Important nonlinear phenomena}
Nonlinear systems are known for exhibiting rich and sophisticated dynamical phenomena which include the dependence of resonance frequency on motion amplitude and the resulting hardening and softening phenomena, co-existence of multiple stable vibration states and jumps between different states due to external perturbations,
existence of unstable branches in the nonlinear frequency response curves (NFRCs) that are interconnected with stable portions at bifurcation points,
activation of multiharmonic responses by a mono-harmonic excitation,
appearance of secondary resonances, namely superharmonic and subharmonic resonances, in addition to the primary resonance, emergence of isolated resonance curves detached from the main branch,
occurrence of internal resonances and exchange of energy between different modes, etc \cite{nayfeh2008}. These dynamical phenomena pose great challenges for the development of robust vibration testing schemes.

%\subsection{Important nonlinear concepts}

In this context, nonlinear normal modes (NNMs), defined either as periodic motions of an undamped and unforced mechanical system \cite{rosenberg1960,rosenberg1962,kerschen2009}, invariant manifolds in phase space \cite{shaw1991,shaw1993} or spectral submanifolds \cite{haller}, have become a central concept in the structural dynamics community. NNMs offer valuable insights into the dynamics of a nonlinear system near resonance \cite{hill2015}. For instance, the NNM theory proved useful for the analysis of slender structures featuring geometrical nonlinearity as, e.g., in \cite{touze2021}. In analogy to the frequency response function of a linear system, the NFRCs are obtained by computing branches of periodic responses of a damped, harmonically-forced system. The backbone curves trace the locus of the resonance peaks, establishing a direct link between NNMs and NFRCs. 

%\subsection{Experimental testing based on control}
One of the first attempts to identify NNMs experimentally exploited a nonlinear extension of the phase lag quadrature criterion \cite{peeters2011,peeters2011_2} which, for a linear system, states that a normal mode is excited when the displacement and harmonic forcing are in quadrature. This is at the root of nonlinear phase resonance testing. Following NNM appropriation, their amplitude-dependent properties can, in turn, be extracted during free decay oscillation according to the invariance property of NNMs. More recently, phase-locked loop testing (PLLT) was introduced to robustify NNM identification \cite{peter2017,scheel2018,denis2018,givois2020,abeloos2022}. PLLT makes use of feedback-controlled excitation to impose a user-specified phase lag between the applied excitation and the measured response.
To this end, the excitation frequency is automatically adjusted until the control objective is met. An NFRC can be obtained by sequentially increasing or decreasing the phase lag at constant forcing amplitude to sweep the resonance peak whereas a backbone curve can be measured by sequentially increasing the forcing amplitude while ensuring phase quadrature conditions.

The vast majority of the studies in the technical literature dealing with experimental modal analysis of nonlinear structures focuses on the characterization of primary resonances. However, a fundamental harmonic excitation can activate fractional and higher-order harmonic responses, which, in turn, can trigger the excitation of secondary resonances. Even if the superharmonic and subharmonic resonances of nonlinear systems
have been widely studied analytically and numerically, see, e.g., \cite{nayfeh2008,parlitz,vizzaccaro2023}, their experimental identification remains a field to be explored. In this context, we note the pioneering works by Chauhan et al. \cite{chauhan1971} and Yamaki et al. \cite{yamaki1981,yamaki1983}. Experimental studies on other physical systems also reported the existence of subharmonic resonance phenomena, such as acoustic emission from gas bubbles \cite{neppiras1969}, nonlinear electrical LC circuits \cite{janssen1984}, a moored ocean structural system \cite{lin1998}, shear layer flow under excitation \cite{husain1995}. Reference \cite{zhang2017} considers the monitoring of subharmonic response components for detecting bolted joint loosening.

The objective of this study is to develop a vibration testing scheme based on PLL feedback control and the concept of a resonant phase lag \cite{volvert2021} for the identification of the secondary resonances of a nonlinear mechanical system.
To this end, PLL is coupled to an adaptive digital filter, a powerful building block which was first incorporated in the control loop in \cite{abeloos2021}. The adaptive filter allows the analyst to perform online Fourier decomposition and, hence, estimate the phase lag of a target harmonic of the response with respect to the harmonic forcing. The resonant phase lag  of the considered secondary resonance is then exploited for designing the appropriate control strategy to identify its NFRC and backbone curve.

The paper is organized as follows. The basic principle of the proposed method is introduced in Section 2. In Section 3, the control scheme is tested on a Duffing oscillator in a virtual experiment. The performance of PLLT for the characterization of folded superharmonic responses and subharmonic isolas is investigated.
In Section 4, a slender beam exhibiting geometrical nonlinearity is employed as the test system in a physical experiment. Conclusions are drawn in the final section.

%==================================================================
\section{Phase-locked loop testing with adaptive filters}
%==================================================================

\begin{figure}[htb!]
\begin{center}
\includegraphics[width=0.45\textwidth]{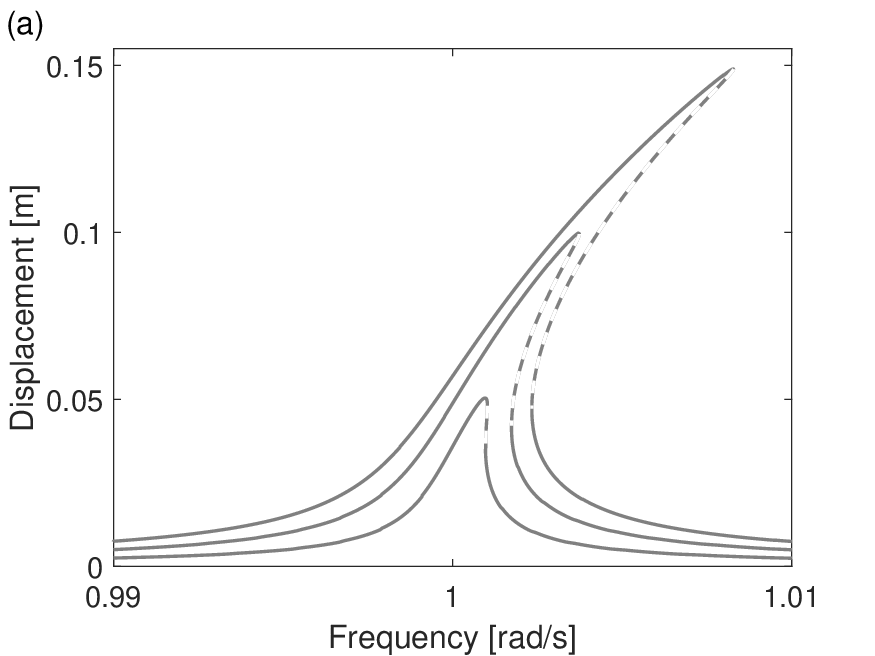} 
\includegraphics[width=0.45\textwidth]{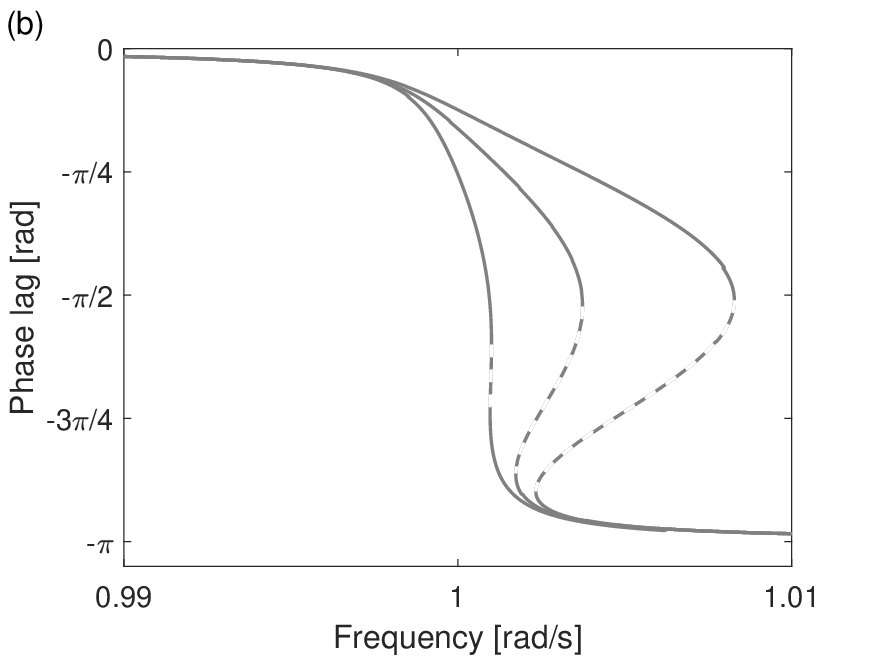}  
\end{center}
\caption{NFRCs of a hardening Duffing oscillator computed at different forcing amplitudes by the harmonic balance method. The unstable responses are indicated by dashed lines. (a) Amplitude and (b) phase lag.}
\label{fig_FRCs_11_In}
\end{figure}

The basic principle of PLLT is to make use of the unique phase variation property of a nonlinear resonance to unfold the NFRCs. 
This is exemplified for the nonlinear primary resonance of a hardening Duffing oscillator excited by a sine forcing function in Figure~\ref{fig_FRCs_11_In}. 
It is seen that the phase lag between the fundamental harmonic of the displacement and the forcing reduces monotonically from 0 to $-\pi$ when increasing the forcing frequency. Thus, there is a unique forcing frequency corresponding to a specific value of the phase lag. In addition, the backbone curve describing the locus of the resonance peaks can be tracked by locking the phase angle at a value of $-\pi/2$ while varying the force amplitude. 

\begin{figure}[htb!]
\begin{center}
\includegraphics[width=0.7\textwidth]{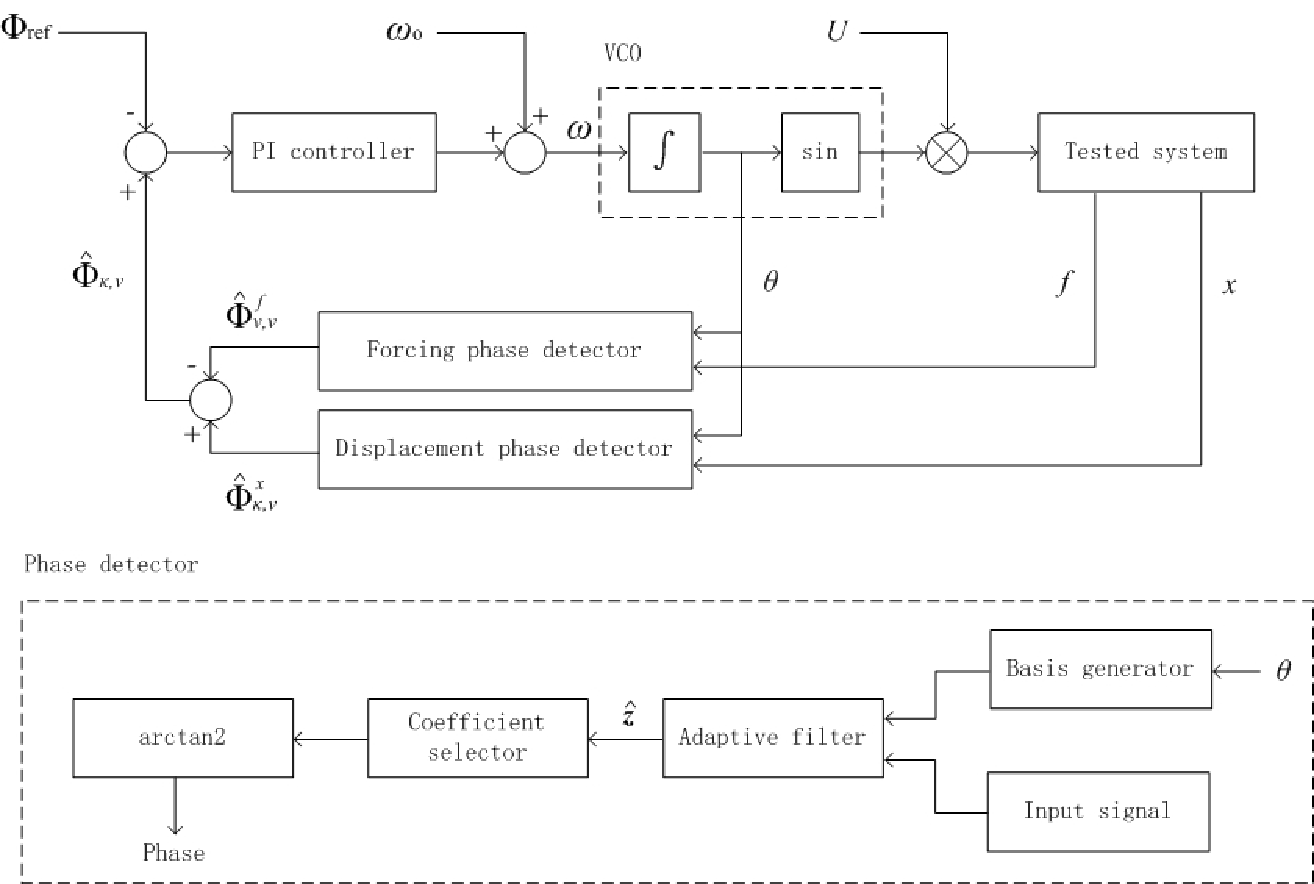}
\end{center}
\caption{PLL scheme with adaptive filters.}
\label{fig:block_diagram}
\end{figure}

The PLL scheme utilized in this work requires no prior knowledge of the tested system and can be easily implemented in a physical experiment. Another benefit is that it can stabilize unstable branches through the action of the controller \cite{denis2018,abeloos2022thesis}.
As schematized in Figure \ref{fig:block_diagram}, it comprises there important building blocks, namely a phase detector, a proportional-integral (PI) controller and a voltage-controlled oscillator (VCO). A closed-loop system is built after the inclusion of the mechanical system to be tested. The system state is prescribed through the imposition of a desired phase lag between the harmonic excitation and a specific harmonic of the response. To this end, the forcing frequency is automatically adjusted by the controller.
The harmonic excitation signal is generated by the VCO module based on the fact that the circular frequency is a phase change per unit  time. The external forcing is thus expressed as
\begin{eqnarray}\label{PLL_force}
f(t)=F\sin{ \left[ \int_0^t\omega(\tau)\mathrm{d}\tau \right]}.
\end{eqnarray}
The instantaneous frequency is determined by the difference between the assigned reference phase $\Phi_\mathrm{ref}$ and the phase lag of the $\kappa$-th harmonic $\Phi_{\kappa,\upsilon}$ obtained by the phase detector and the settled PI controller
\begin{eqnarray}\label{PLL_freq}
\omega(t) = \omega_\mathrm{o} + K_P \left[\Phi_{\kappa,\upsilon}(t)-\Phi_\mathrm{ref} \right] + K_I\int_0^t \left[ \Phi_{\kappa,\upsilon}(\tau) -\Phi_\mathrm{ref} \right] \mathrm{d}\tau.
\end{eqnarray}
Here, $\omega_\mathrm{o}$ is the open-loop forcing frequency, often set to be close to the target resonance.
$K_P$ and $K_I$ are the proportional and integral gains of the controller, respectively.
They are employed to achieve fast convergence to the target phase angle and eliminate the steady-state errors in the feedback loop. Alternatively, Eq. (\ref{PLL_freq}) can be rewritten into a set of differential equations \cite{denis2018}. A differential gain can be introduced to further accelerate the convergence, but it is not considered in this work for simplicity.


The measured periodic response is written in terms of the truncated Fourier expansion
\begin{eqnarray}\label{sin_expansion}
x(t)=\dfrac{c_{0,\upsilon}}{\sqrt{2}} + \sum_{\kappa=1}^N \left[ s_{\kappa,\upsilon} \sin{\left( \dfrac{\kappa \omega}{\upsilon}t \right)} + c_{\kappa,\upsilon} \cos{\left( \dfrac{\kappa \omega}{\upsilon}t \right)} \right].
\end{eqnarray}
where $\kappa$ is the number of harmonics and $\upsilon$ is a positive integer introduced to account for subharmonics. The fundamental harmonic excitation can activate multiharmonic responses which can consequently trigger the excitation of secondary resonances, namely $\kappa:1$ superharmonic resonances, $1:\upsilon$ subharmonic resonances or $k:\upsilon$ ultrasubharmonic resonances \cite{volvert2021}, when the frequency of the $\kappa$-th harmonic $\omega_{\kappa,\upsilon} = \kappa \omega / \upsilon$ corresponds to that of 1:1 primary resonance.
The expansion (\ref{sin_expansion}) can be recast into a more compact form
\begin{eqnarray}\label{Mat_expansion}
x(t) = \textbf{Q}(t) \textbf{z},
\end{eqnarray}
where $\textbf{Q}(t)$ is a row vector collecting the time-dependent trigonometric terms
\begin{eqnarray}\label{Mat_Q}
\textbf{Q}(t) = \left[ \dfrac{1}{\sqrt{2}} \; \sin{\left( \dfrac{\omega}{\upsilon}t \right)} \; \cos{\left( \dfrac{\omega}{\upsilon}t \right)} \; ... \; \sin{\left( \dfrac{N\omega}{\upsilon}t \right)} \; \cos{\left( \dfrac{N\omega}{\upsilon}t \right)}\right],
\end{eqnarray}
and $\textbf{z}$ is a vector containing the associated Fourier coefficients
\begin{eqnarray}\label{Mat_z}
\textbf{z} = \left[ c_{0,\upsilon} \; s_{1,\upsilon} \; c_{1,\upsilon} \; ... \; s_{N,\upsilon} \; c_{N,\upsilon} \right]^\mathrm{T},
\end{eqnarray}
In this work, the amplitude and phase lag of the $\kappa$-th harmonic component are determined as
\begin{eqnarray}\label{kth_harm}
x_{\kappa,\upsilon}(t) =  A_{\kappa,\upsilon}^x \sin{\left( \dfrac{\kappa \omega}{\upsilon}t + \Phi_{\kappa,\upsilon}^x \right)} ,
\end{eqnarray}
where 
\begin{eqnarray}\label{amplitude}
A_{\kappa,\upsilon}^x =  \sqrt{c_{\kappa,\upsilon}^2 + s_{\kappa,\upsilon}^2}, 
\end{eqnarray}
and 
\begin{eqnarray}\label{phase}
\Phi_{\kappa,\upsilon}^x =  \mathrm{arctan2} ( c_{\kappa,\upsilon}, s_{\kappa,\upsilon}).
\end{eqnarray}

To develop a control-based scheme that can effectively characterize secondary resonances, the phase of the non-fundamental harmonic of interest, i.e., $\kappa \ne \upsilon$, must be accurately estimated. This is why we resort to adaptive digital filters for performing online Fourier decomposition \cite{Haykin}. Indeed, reference \cite{abeloos2022thesis} demonstrated that adaptive filtering presents distinct advantages over the conventional synchronous demodulation techniques based on low-pass filters. Adaptive filtering synthesizes a signal $\hat{x}(t)$ by considering a time-dependent basis weighted by a set of coefficients, such that it approximates the measured response $x(t)$
\begin{eqnarray}\label{AF_approx}
\hat{x}(t) = \textbf{Q}(t) \hat{\textbf{z}} \approx x(t),
\end{eqnarray}
The row vector $\textbf{Q}(t)$ containing the sine and cosine functions now becomes the basis in adaptive filtering, and the weight coefficients $\hat{\textbf{z}}$ are combined in the vector
\begin{eqnarray}\label{AF_weight}
\hat{\textbf{z}} = \left[ \hat{z}_{0} \; \hat{z}_{1s} \; \hat{z}_{1c} \; ... \; \hat{z}_{Ns} \;\hat{z}_{Nc} \right]^\mathrm{T}.
\end{eqnarray}
These weights can be interpreted as an approximation of the Fourier coefficients with $\hat{\textbf{z}} \approx \textbf{z}$. 
They are updated online at discrete time steps using the least mean squares (LMS) algorithm, which is more cost-effective as compared to other alternative algorithms \cite{abeloos2021}.
The weights at the instant $t_{i+1}$ are updated according to the gradient of the mean squared error at $t_i$  
\begin{eqnarray}\label{AF_updating}
\hat{\textbf{z}}(t_{i+1}) = \hat{\textbf{z}}(t_i) + \mu \epsilon(t_i) \textbf{Q}^\mathrm{T}(t_i),
\end{eqnarray}
where $\mu$ is the step size factor specifying the convergence and $\epsilon$ is the synthesis error estimated using
\begin{eqnarray}\label{AF_error}
\epsilon(t_i) = x(t_i) - \hat{x}(t_i) = x(t_i) - \textbf{Q}(t_i) \hat{\textbf{z}}(t_i),
\end{eqnarray}
The phase of the $\kappa$-th harmonic of the displacement can then be extracted according to
\begin{eqnarray}\label{est_phase}
\hat{\Phi}_{\kappa,\upsilon}^x =  \mathrm{arctan2} ( \hat{z}_{\kappa c}, \hat{z}_{\kappa s}) . %\approx \Phi_{\kappa,\upsilon}
\end{eqnarray}
The phase of the fundamental forcing component at $\kappa=\upsilon$ denoted $\hat{\Phi}_{\upsilon,\upsilon}^f$ can be decomposed in a similar manner, so that the phase lag of the $\kappa$-th harmonic of the displacement with respect to the forcing can be estimated using the relation 
\begin{eqnarray}\label{est_phase_lag}
\hat{\Phi}_{\kappa,\upsilon} =  \hat{\Phi}_{\kappa,\upsilon}^x-\hat{\Phi}_{\upsilon,\upsilon}^f \approx \Phi_{\kappa,\upsilon}.
\end{eqnarray}
When it comes to practical implementation, the phase angles of the trigonometric terms of the basis in (\ref{AF_approx}) are estimated through the integration of the instantaneous frequency with respect to time, to ensure a smooth phase variation. 

Eventually, the concept of a resonant phase lag introduced in \cite{volvert2021} can be exploited for the identification of the NFRC and backbone of the resonance of interest. For instance, for a Duffing oscillator subjected to a sine forcing function, it is shown that the points where the phase lag is $-\pi/2$ ($\kappa$ and $\upsilon$ odd) or $-3\pi/4\upsilon$ ($\kappa$ or $\upsilon$ even) define the locus of phase resonance of the $\kappa:\upsilon$ resonance.
An important remark is that the phase lags used in this study are different from those in \cite{volvert2021} owing to difference in the sign conventions used to define phase lag.
The resonant phase lags of other oscillators are given in \cite{volvertthesis}.

%==================================================================
\section{Numerical demonstration using a Duffing oscillator}
%==================================================================

A Duffing oscillator under sine forcing governed by the equation of motion
\begin{eqnarray}\label{EoM_Duffing}
m\ddot{x}(t) + c\dot{x}(t) + kx(t) + k_{nl}x^3(t) = F\sin{\omega t},
\end{eqnarray}
is considered in this section. $m$, $c$, $k$ and $k_{nl}$ are the mass, damping, linear and nonlinear stiffness coefficients, respectively. They are chosen to be equal to $m=1$ kg, $c=0.001$ kg/s, $k=1$ N/m and $k_{nl}=1$ N/m$^3$. The natural frequency of the underlying linear system is $\omega_{l}=\sqrt{k/m}$.

The dynamical model integrating PLL control and the Duffing oscillator was implemented in the Matlab/Simulink environment. The fixed-step ordinary differential equation solver was employed for carrying out the time-domain simulations. 
The parameters of the PLL controller and of the adaptive filter are tabulated in Table~\ref{tab:parameters} where $f_s$ is the sampling frequency.

\begin{table}[htb!]
\centering
\begin{tabular}{c|c|c|c|c}
Target resonance  & $K_P$ [1/s] & $K_I$ [1/s$^2$] & $f_s$ [1/s] & $\mu$ [-] \\\hline
1:1         & $1.0$ & $5 \times 10^{-3}$ & $2 \times 10^2$ & $10^{-4}$\\
3:1 and 2:1 & $0.1$ & $10^{-4}$ & $2 \times 10^2$ & $10^{-4}$\\
1:3 and 1:2 & $2.0$ & $10^{-2}$ & $10^3$ & $10^{-4}$\\
\end{tabular}
\caption{Parameters used in the virtual experiment.}
\label{tab:parameters}
\end{table}

\subsection{Primary resonance}

\begin{figure}[htb!]
\begin{center}
\includegraphics[width=0.45\textwidth]{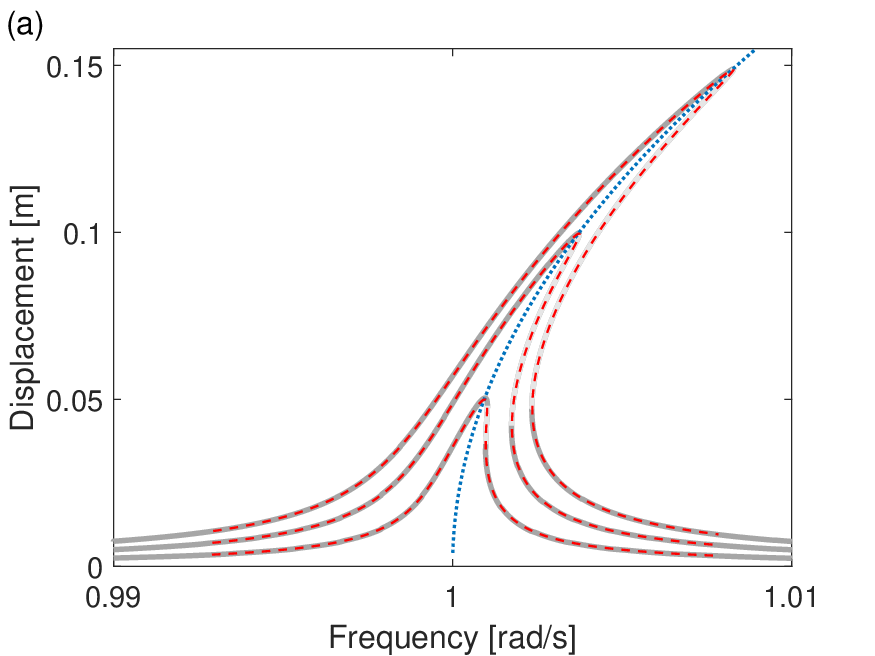} 
\includegraphics[width=0.45\textwidth]{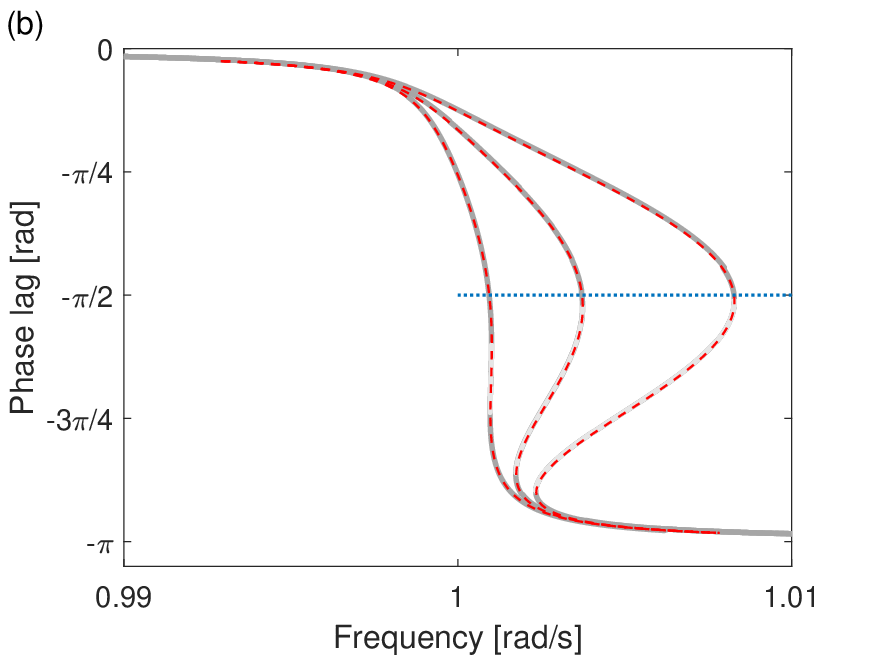}  
\end{center}
\caption{NFRCs (\tpf{red}) and the backbone curve (\tppp{NavyBlue}) of the primary resonance identified by PLLT. The reference solution is provided by HBM in gray; light gray corresponds to unstable responses. (a) Amplitude and (b) phase lag.}
\label{fig_FRCs_11}
\end{figure}

The NFRC and backbone of the primary resonance  were identified using PLLT at three different forcing amplitudes $F$, namely $0.00005$, $0.0001$ and $0.00015$ N. A reference solution was also computed using the harmonic balance method (HBM) \cite{detroux2015} with 14 harmonics. The results are plotted in Figure~\ref{fig_FRCs_11}. Clearly, PLLT can trace out very accurately both the stable and unstable branches of the NFRC. In addition, the backbone curve corresponding to nonlinear phase resonance conditions can also be tracked accurately by locking the phase lag at $-\pi/2$ and varying the forcing amplitude. 
%An important remark is that the phase lags used in this study are different from those given in \cite{volvert2021} owing to the fact that a sine forcing function, and not a cosine forcing function, is considered herein.

\begin{figure}[htb!]
\begin{center}
\includegraphics[width=0.45\textwidth]{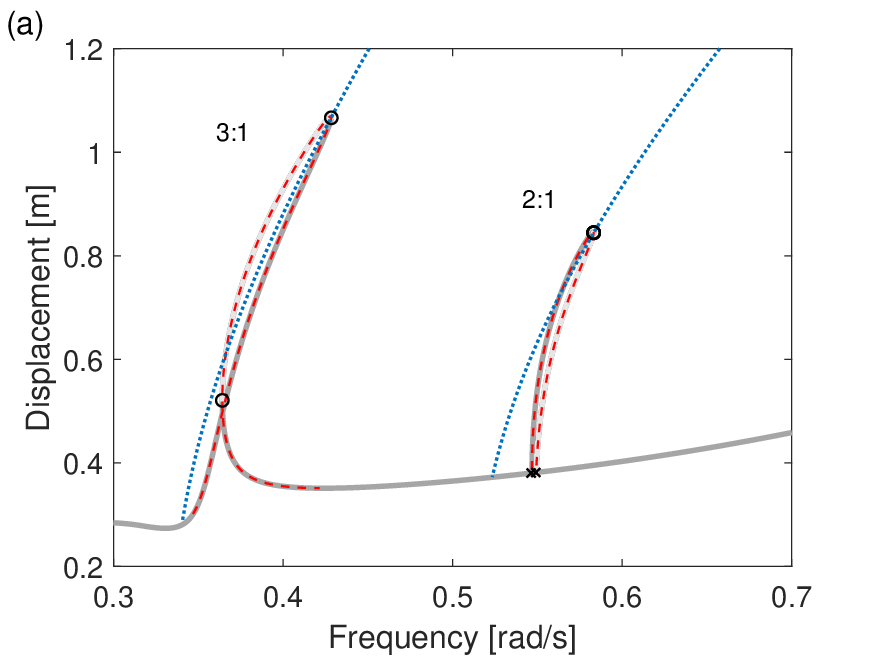} 
\includegraphics[width=0.45\textwidth]{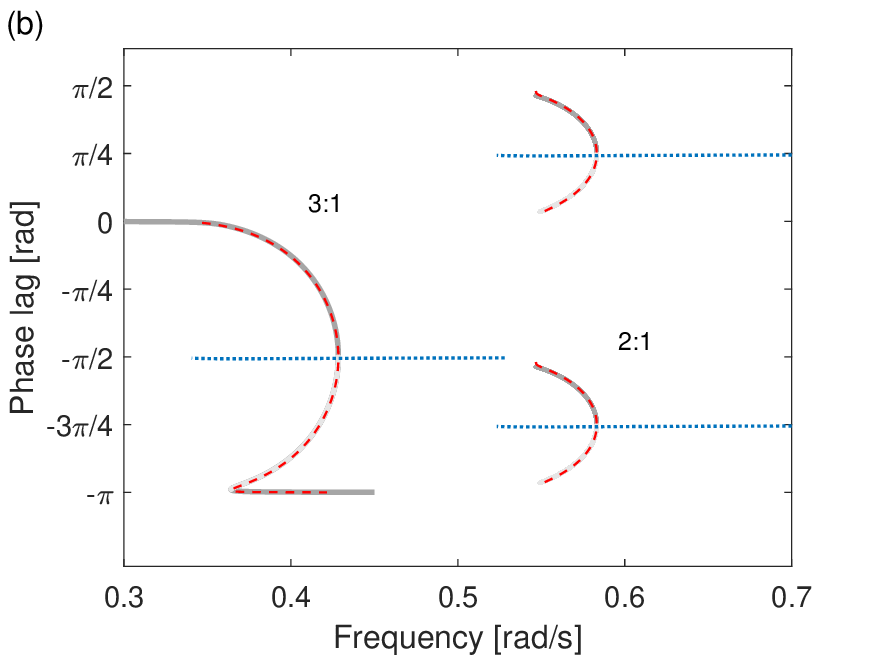}  
\end{center}
\caption{NFRCs (\tpf{red}) at $F=0.3$N and backbone curves (\tppp{NavyBlue}) of the 3:1 and 2:1 resonances identified by PLLT. The reference solution is provided by HBM in gray; light gray corresponds to unstable responses. The fold and branch-point bifurcations are marked with circles and crosses, respectively. (a) Amplitude and (b) phase lag.}
\label{fig_FRCs_3121_f03}
\end{figure}
\begin{figure}[htb!]
\begin{center}
\includegraphics[width=0.45\textwidth]{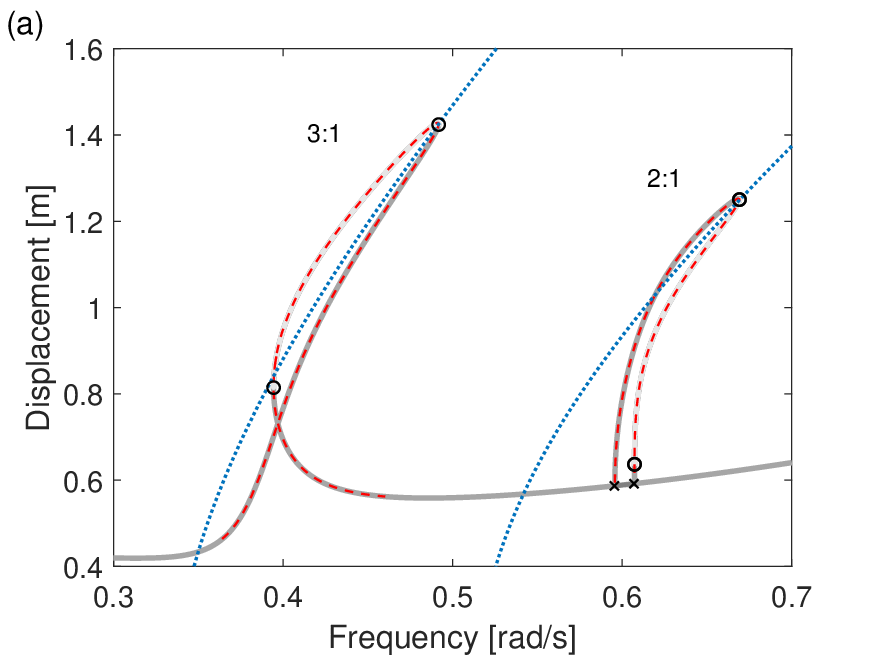} 
\includegraphics[width=0.45\textwidth]{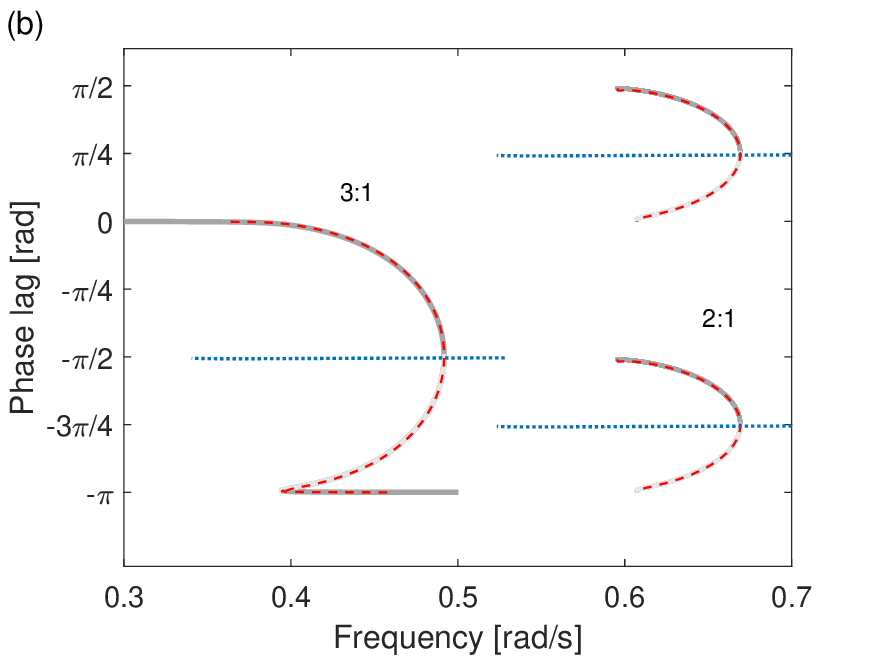}  
\end{center}
\caption{NFRCs (\tpf{red}) at $F=0.5$N and backbone curves (\tppp{NavyBlue}) of the 3:1 and 2:1 resonances identified by PLLT. The reference solution is provided by HBM in gray; light gray corresponds to unstable responses. The fold and branch-point bifurcations are marked with circles and crosses, respectively. (a) Amplitude and (b) phase lag.}
\label{fig_FRCs_3121_f05}
\end{figure}

\subsection{Superharmonic resonances}

As shown in \cite{volvert2021} for a hardening Duffing oscillator and in Figures \ref{fig_FRCs_3121_f03} and \ref{fig_FRCs_3121_f05} of this paper, the phase lag $\Phi_{\kappa,1}$ of the $\kappa$:1 superharmonic resonances evolves in a similar manner compared to the phase lag of the primary resonance. Specifically, the phase lag decreases
monotonically when increasing the forcing frequency, a property which is not fulfilled by the fundamental harmonic $\Phi_{1,1}$ of the superharmonic resonance. Again, this paves the way for a simple continuation scheme during which the phase lag is sequentially decreased. Furthermore, around the superharmonic resonance, the superharmonic component $x_\kappa$ makes a significant contribution to the total response $x$, so that $\Phi_{\kappa,1}$ can be accurately extracted. 

The identification of the 3:1 and 2:1 superharmonic resonances was achieved using PLLT by letting the phase lags $\Phi_{3,1}$ and $\Phi_{2,1}$ vary between 0 and $-\pi$ or between $-\pi/2$ and $-\pi$, respectively. The corresponding phase resonances occur at $-\pi/2$ and $-3\pi/4$, respectively. Figures \ref{fig_FRCs_3121_f03} and \ref{fig_FRCs_3121_f05} present the resulting NFRCs and backbone curves at $F=0.3$N and $F=0.5$N, respectively. An excellent identification of the two superharmonic resonances was obtained using PLLT. For the 2:1 resonance, a second solution was also found to exist between 0 and $\pi/2$, as discussed in \cite{volvert2021}. Other odd and even superharmonic resonances can be characterized in a similar manner.

\begin{figure}[htb!]
\begin{center}
\includegraphics[width=0.45\textwidth]{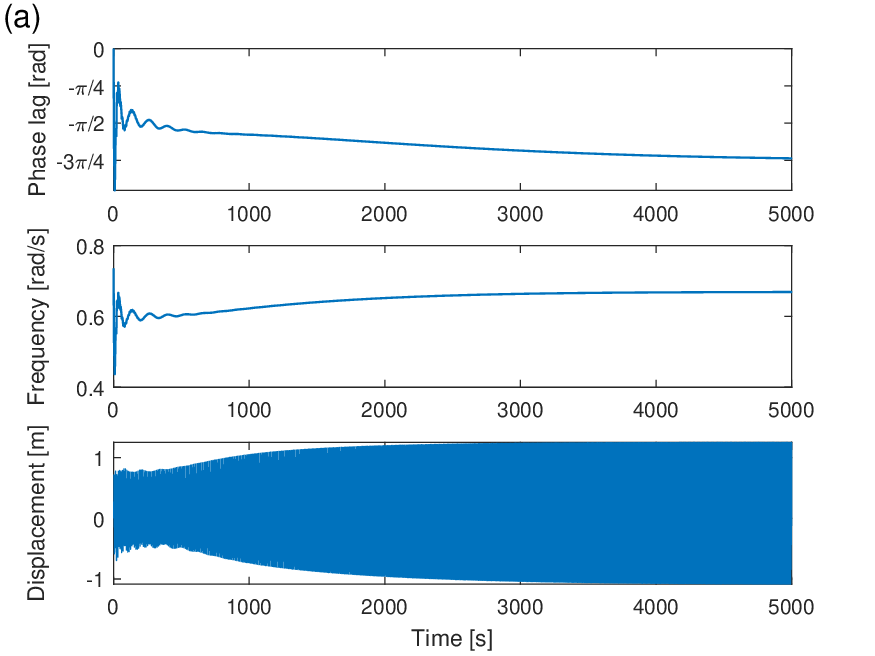}
\includegraphics[width=0.45\textwidth]{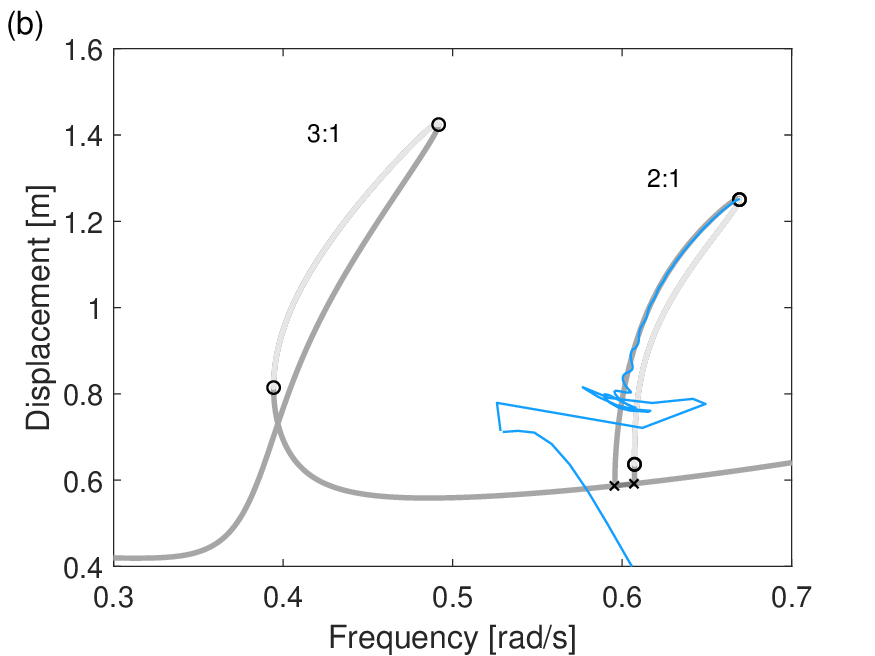}
\end{center}
\caption{Transition from the rest position to the 2:1 superharmonic resonance: $F=0.5$ N, $\omega_\mathrm{o}=0.5$ and $\Phi_\mathrm{ref}=-3\pi/4$. (a) Time evolution of the phase lag, instantaneous frequency and displacement and (b) trajectory in the displacement-frequency plane.}
\label{fig_time_21}
\end{figure}

The calculation of the 3:1 resonance is facilitated by the fact that it appears in the direct continuation of the main branch of the NFRC. This is not the case for the 2:1 resonance which bifurcates out of the main branch. As a result, HBM necessitates either a branch switching strategy or an initial guess to calculate the 2:1 branch. Interestingly, this is not necessary for the proposed PLLT scheme, because the 2:1 branch can be reached from the rest position, i.e., starting with trivial initial conditions $x_0=\dot{x}_0=0$. As illustrated in Figure \ref{fig_time_21}, after a transient period during which the phase lag converges toward $-3\pi/4$, the system's motion reaches the 2:1 branch. We note that $\omega_0$ shopuld be chosen close enough from the frequency range in which the resonance exists.

\begin{figure}[htb!]
\begin{center}
\includegraphics[width=0.45\textwidth]{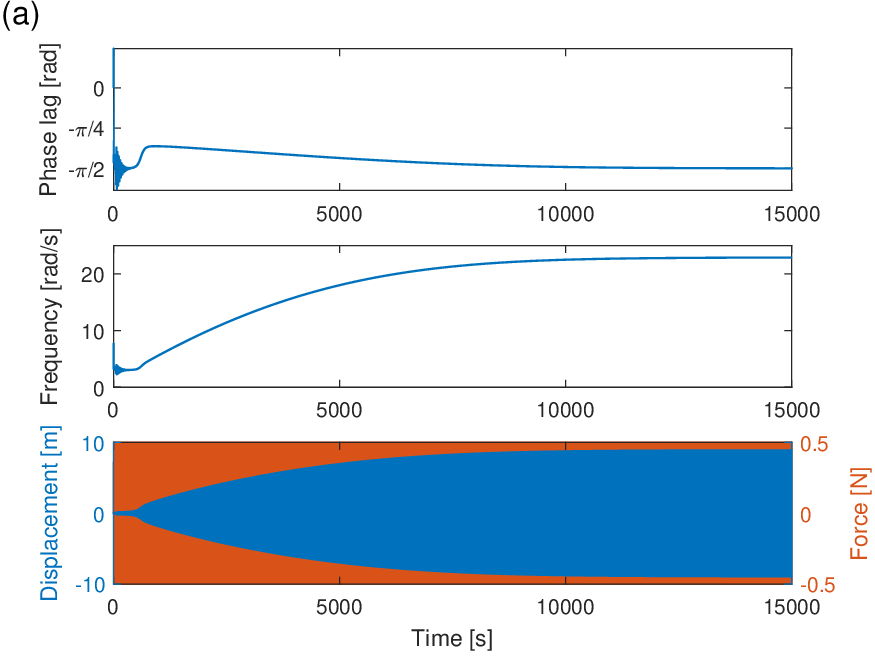}
\includegraphics[width=0.45\textwidth]{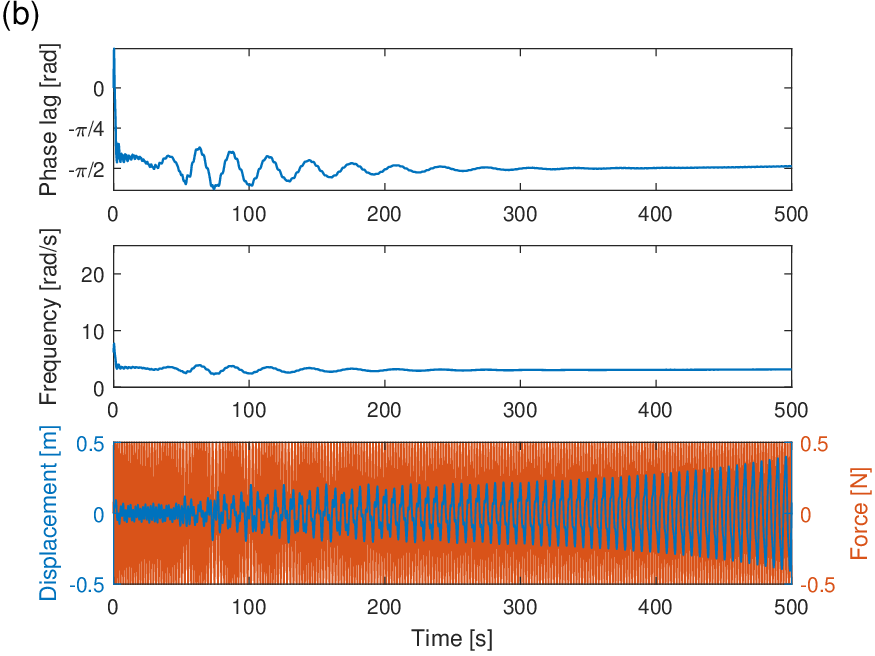}
\end{center}
\begin{center}
\includegraphics[width=0.45\textwidth]{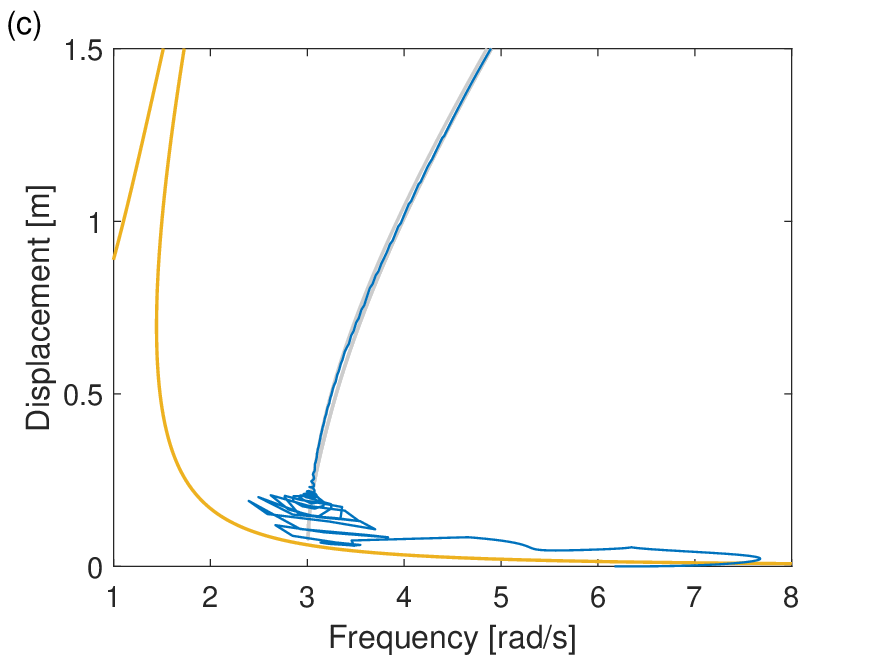}
\includegraphics[width=0.45\textwidth]{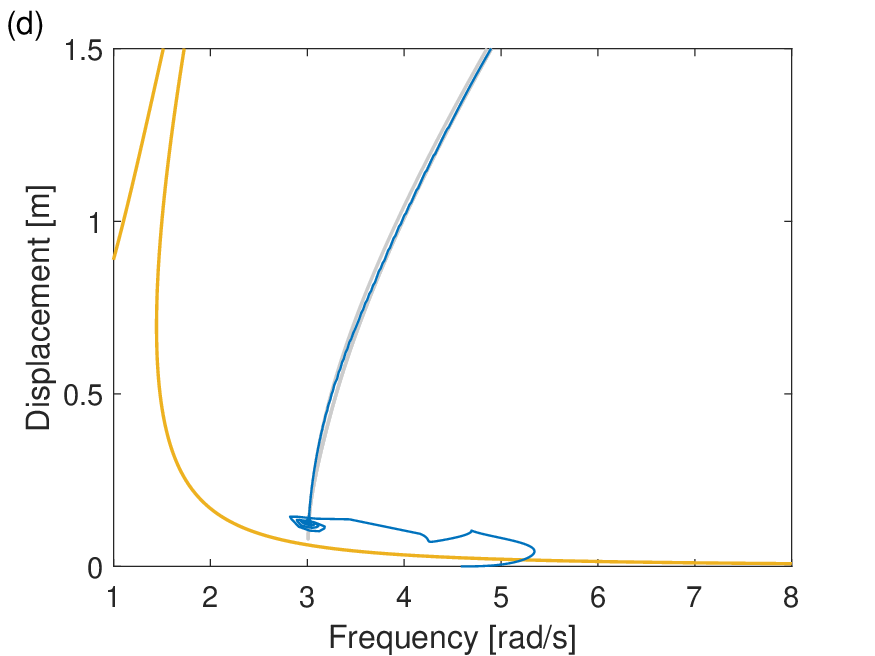}
\end{center}
\caption{Transition from the rest position to the upper resonance point of the 1:3 subharmonic resonance: $F=0.5$ N, $\omega_\mathrm{o}=3$ and $\Phi_\mathrm{ref}=-\pi/2$. (a-b) Time evolution of the phase lag, instantaneous frequency and displacement: (a) complete time interval and (b) close-up on the initial times; (c-d) Trajectory in the displacement-frequency plane: (c) $K_I=2$ and $K_P=0.01$ and (d) $K_I=1$ $K_P=0.01$. The yellow branch corresponds to the main branch of the NFRC.}
\label{fig_time_13}
\end{figure}
\begin{figure}[htb!]
\begin{center}
\includegraphics[width=0.45\textwidth]{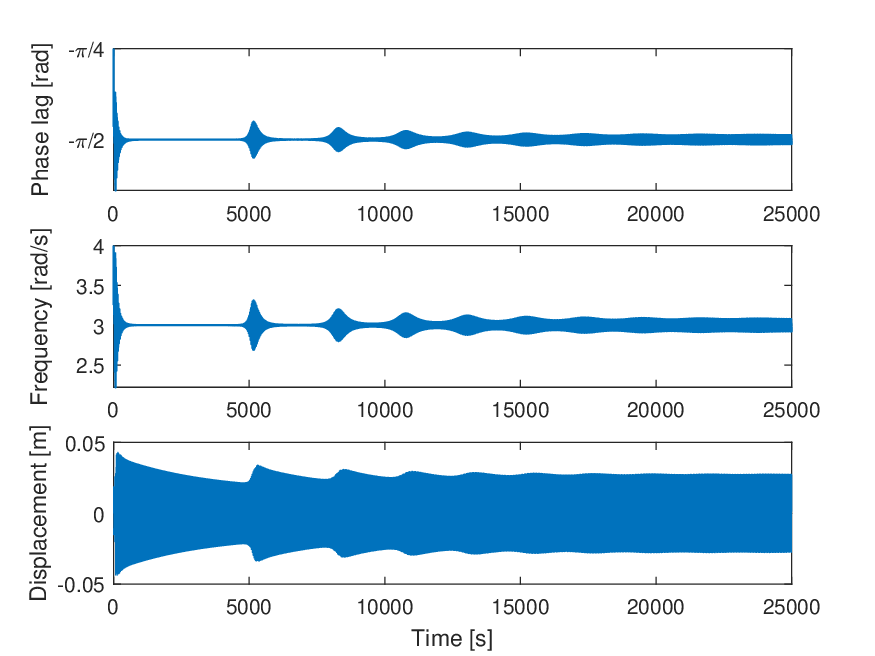}
\end{center}
\caption{Unsuccessful transition from the rest position to the 1:3 subharmonic resonance: $F=0.1$N, $\omega_\mathrm{o}=3$ and $\Phi_\mathrm{ref}=-\pi/2$.}
\label{fig_time_13_p01}
\end{figure}
\begin{figure}[htb!]
\begin{center}
\includegraphics[width=0.45\textwidth]{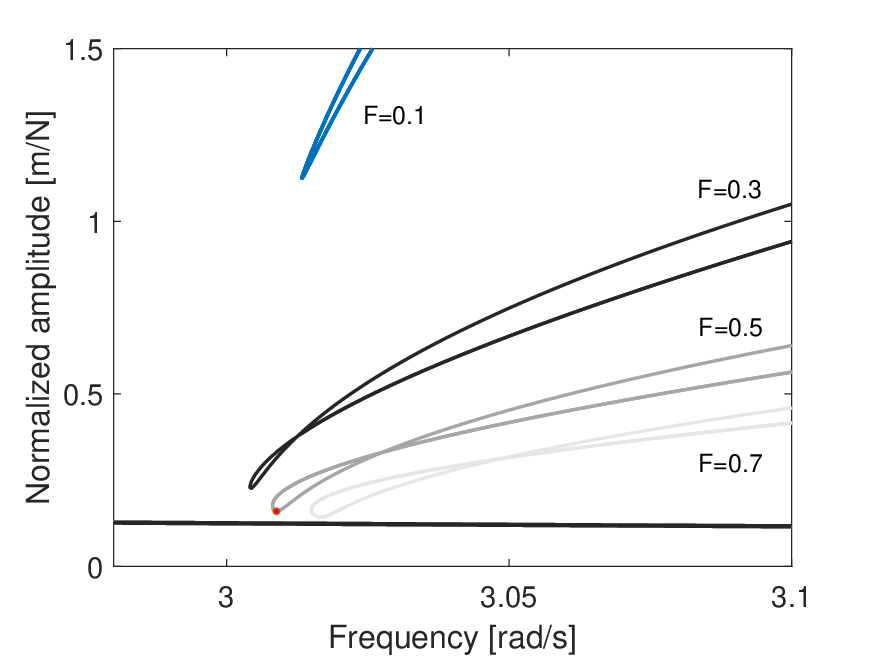}
\end{center}
\caption{Normalized frequency responses $X/F$ of 1:3 resonances computed by HBM. }
\label{fig_XF_13}
\end{figure}
\begin{figure}[htb!]
\begin{center}
\includegraphics[width=0.45\textwidth]{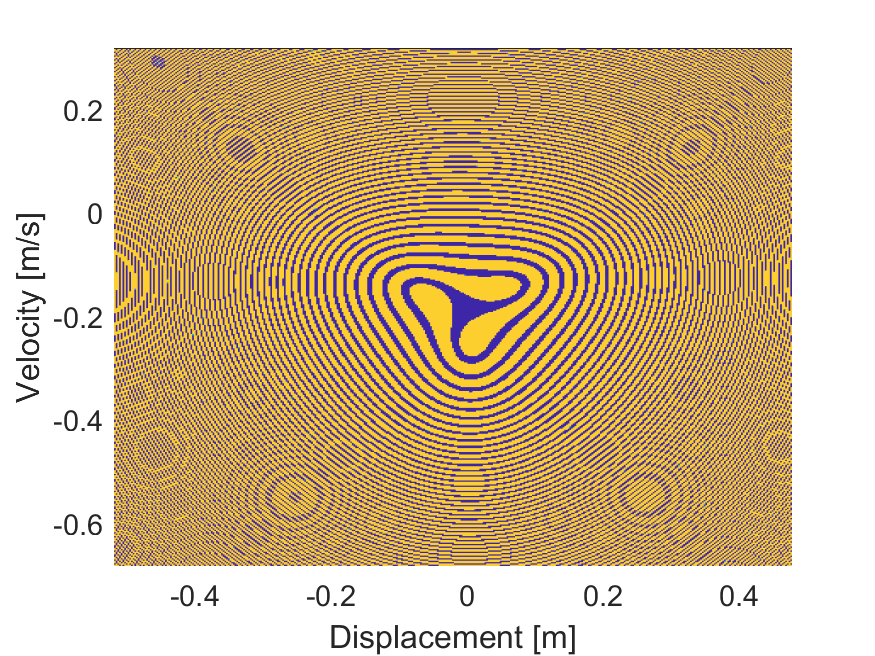}
\end{center}
\caption{Basins of attraction near the red point in Figure \ref{fig_XF_13} for $F=0.5$. Blue and yellow colors correspond to the periodic responses on the main branch of the NFRC and on the 1:3 isola, respectively.}
\label{fig_Basin_of_Attraction}
\end{figure}

\subsection{Subharmonic resonances}

Unlike superharmonic resonances, subharmonic resonances take the form of isolated branches of periodic solutions detached from the main branch \cite{nayfeh2008}. This is why they are more difficult to calculate during numerical continuation and can also be missed when considering classical sine sweep testing. These resonances only exist beyond a specific threshold in forcing amplitude and appear near multiples of the primary resonance frequency.

\begin{figure}[htb!]
\begin{center}
\includegraphics[width=0.45\textwidth]{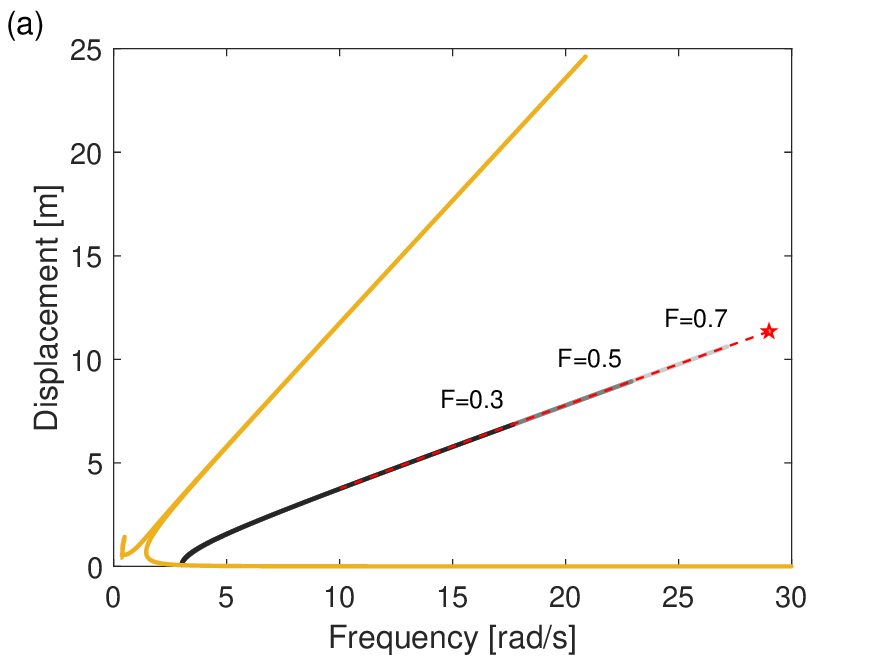} 
\includegraphics[width=0.45\textwidth]{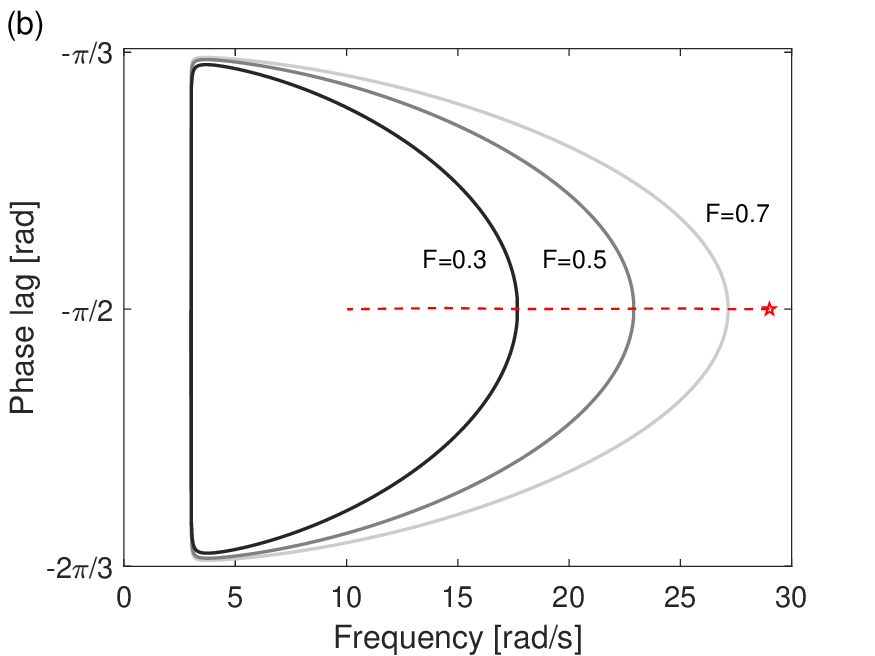}  
\end{center}
\caption{Backbone curve (\tpf{red}) of the 1:3 subharmonic resonance identified by PLLT. The starting point is the upper resonance point at $F=0.8$N indicated by a star marker. The subharmonic branches (gray) are computed by HBM. The main branch of the NFRC at $F=0.5$ N is represented in yellow. (a) Amplitude and (b) phase lag.}
\label{fig_BBs_13}
\end{figure}

As discussed in \cite{volvert2021} and shown in Figure \ref{fig_BBs_13}(b), the phase lag $\Phi_{1,\upsilon}$ of the 1:$\upsilon$ subharmonic resonances presents a more complex topology owing to the closed nature of an isola. However, when either the lower or the upper branch of the isola is considered, there is still a 1:1 relationship between the phase lag and the forcing frequency. For the 1:3 and 1:2 subharmonic resonances, the resonant phase lag is $\Phi_\mathrm{ref}=-\pi/2$ and $\Phi_\mathrm{ref}=-3\pi/8$, respectively. Unlike superharmonic resonances,  
each subharmonic branch possesses two points featuring the resonant phase lag, i.e., one at the lowest extremity of the isola and the other one at the highest extremity \cite{volvert2021}.  

As for the 2:1 resonance, a very interesting feature of PLLT is that subharmonic resonances can be reached without any guess or user interaction. This is illustrated for the 1:3 subharmonic resonance in Figure \ref{fig_time_13} for $F=0.5$N. PLLT starts from the rest position $x_0=\dot{x}_0=0$ and imposes the resonant phase lag $\Phi_\mathrm{ref}=-\pi/2$. A so-called {\it state transfer} process to the upper resonance point of the 1:3 isola then occurs. There are mainly three stages in this transient process:
\begin{itemize}
    \item[i] An initial overshooting, a typical phenomenon when using a PID controller, results in fast oscillations of the instantaneous frequency around $\omega=3$. This overshooting depends on the controller parameters $K_P$ and $K_I$. For instance, Figure \ref{fig_time_13}(d) shows that a smaller $K_P$ results in less severe oscillations. These oscillations provide the trigger necessary to the jump onto the isola. Once the oscillations are stabilized, the system state is attracted in the vicinity of the low-amplitude resonance point of the isola.
    \item[ii] A smooth transfer along the subharmonic branch toward the high-amplitude resonant point takes place. The fact that the PLL controller has a general tendency to reach the upper resonance point can be explained by the fact that there is an extremely rapid change in the phase lag around the lower amplitude resonance point (see around $\omega \approx 3$ in Figure \ref{fig_BBs_13}), which complicates convergence toward this point. 
    \item[iii] When the measured phase lag converges to the reference phase lag, the upper resonance point of the subharmonic resonance is reached.
\end{itemize}

Such an {\it automatic} branch switching process could be achieved at other force amplitudes or with reference phase lags different from the resonant phase lag. However, it also happened that the transition to the isola could not be achieved. This is the case for $F=0.1$N in Figure \ref{fig_time_13_p01} where the forcing frequency of the closed-loop system and the phase lag undergo oscillations that cannot be stabilized. Figure \ref{fig_XF_13} which displays the response amplitude normalized by the force amplitude $X/F$ around $\omega=3$ reveals that the distance between the main branch of the NFRC and the 1:3 subharmonic isola increases when the forcing level decreases, hence complicating the transfer process. Further confirmation is given by comparing the basins of attraction of the periodic solution on the main branch and of the periodic solution on the isola for $F=0.5$N in Figure \ref{fig_Basin_of_Attraction}. Clearly, the subharmonic response possesses a large basin of attraction; its area represents $64.9\%$ of the total area. This ratio can increase up to $84.4\%$ when $F=0.7$N whereas it is close to $0\%$ when $F=0.1$N.

Once the transfer process to the upper point of the 1:3 subharmonic isolas has been achieved, the backbone can be identified using PLLT with $\Phi_{\mathrm{ref}}=-\pi/2$. Figure~\ref{fig_BBs_13} shows the outcome of PLLT when the identification is initiated for $F=0.8$N and the forcing level is subsequently decreased.

%\begin{figure}[htb!]
%\begin{center}
%\includegraphics[width=0.45\textwidth]{figures/zz_report_13_f01_time.eps}
%\includegraphics[width=0.45\textwidth]{figures/zz_report_13_f05_time.eps}
%\end{center}
%\caption{Forcing signal in first two periods and subharmonic displacement response. Parameters used in the left figure are $F=0.1$, $\omega=3.0133$, $x_0=-0.1081$, $\dot{x}_0=-0.0342$. Those in the right one are $F=0.5$, $\omega=3.0081$, $x_0=-0.0361$, $\dot{x}_0=-0.1748$.}
%\label{fig_harmonics_2T}
%\end{figure}
	
%\begin{table}[htb!]
%\centering
%\begin{tabular}{c|c|c|c}
%Mode & $F=0.3$ & $F=0.5$ & $F=0.7$\\\hline
%1:3 & $30.8\%$ & $64.9\%$ & $84.4\%$\\
%1:2 & $0\%$ & $0\%$ & $0\%$ \\
%\end{tabular}
%\caption{Ratio of area occupied subharmonic responses in the basin of attraction.}
%\label{tab:basin_of_attraction}
%\end{table}

\begin{figure}[htb!]
\begin{center}
\includegraphics[width=0.45\textwidth]{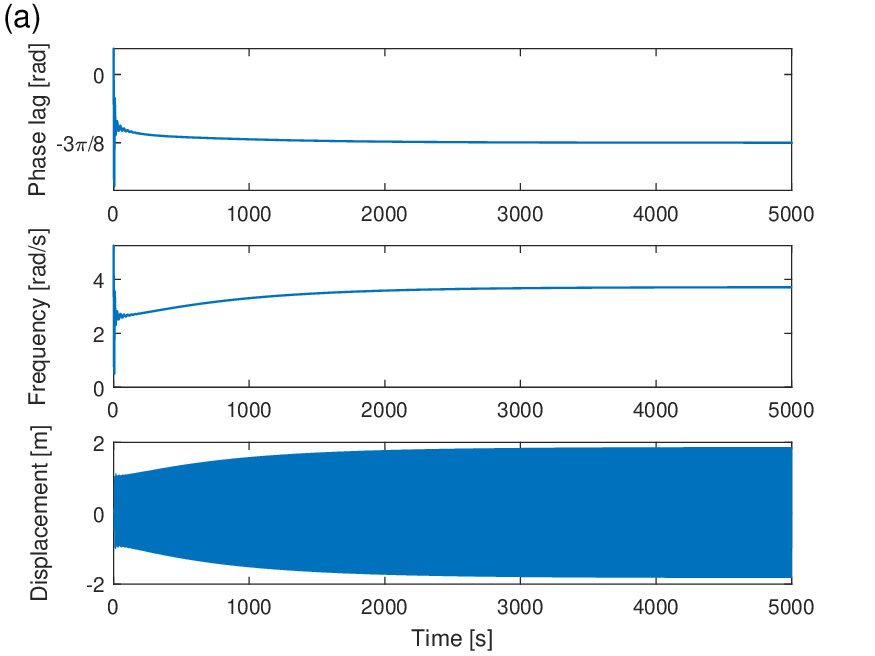}
\includegraphics[width=0.45\textwidth]{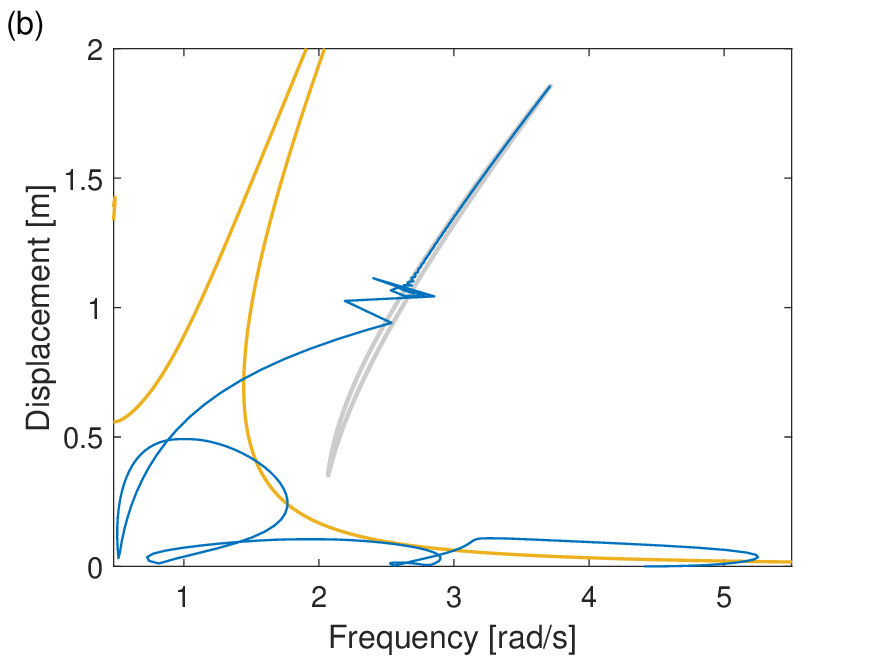}
\end{center}
\caption{Transition from the rest position to the upper resonance point of the 1:2 subharmonic resonance: $F=0.5$ N, $\omega_\mathrm{o}=2$ and $\Phi_\mathrm{ref}=-3\pi/8$. (a) Time evolution of the phase lag, instantaneous frequency and displacement and (b) trajectory in the displacement-frequency plane. The yellow branch corresponds to the main branch of the NFRC.}
\label{fig_time_12}
\end{figure}

\begin{figure}[htb!]
\begin{center}
\includegraphics[width=0.45\textwidth]{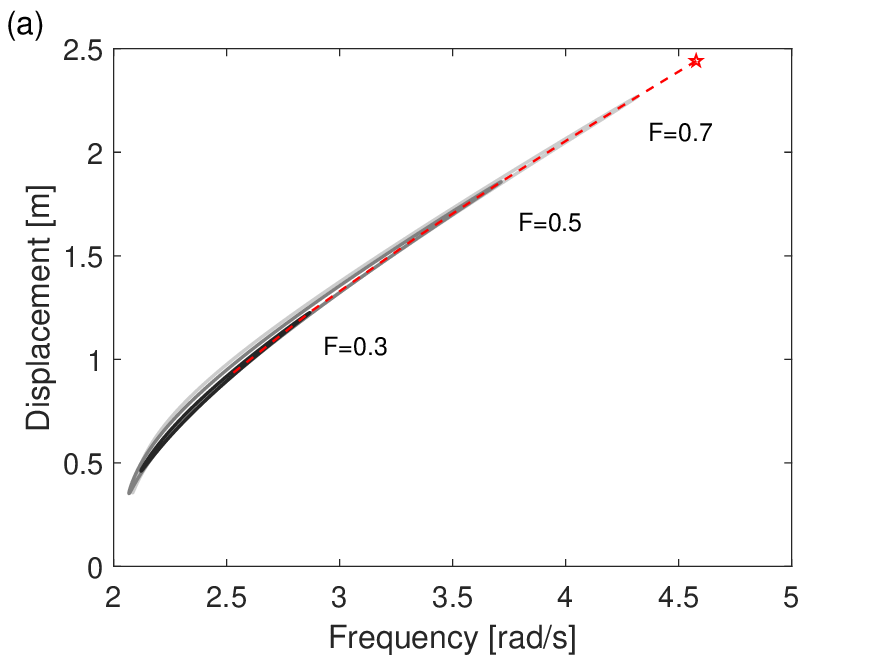}
\includegraphics[width=0.45\textwidth]{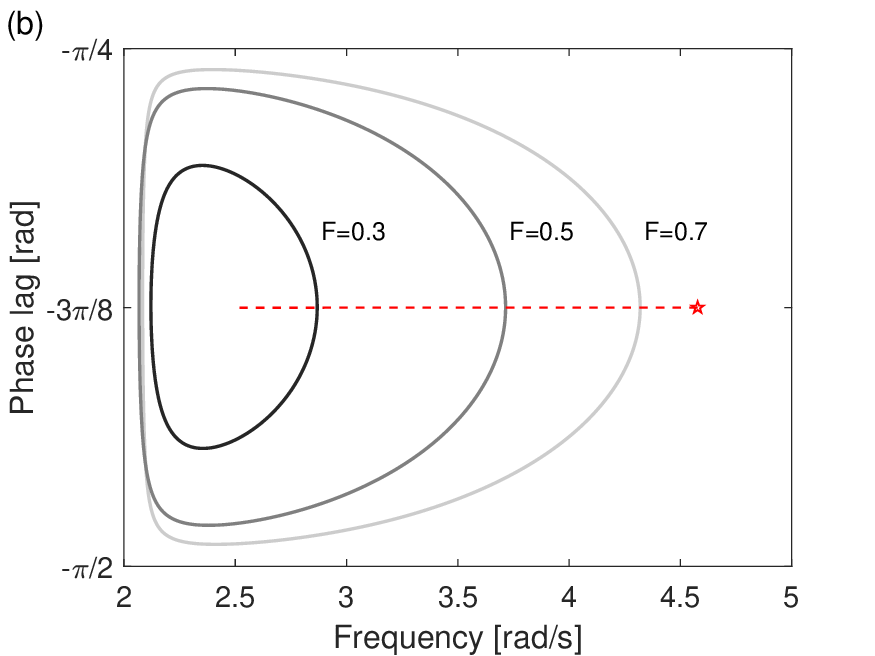}
\end{center}
\caption{Backbone curve (\tpf{red}) of the 1:2 subharmonic resonance identified by PLLT. The starting point is the upper resonance point at $F=0.8$N indicated by a star marker. The subharmonic branches (gray) are computed by HBM. The yellow branch corresponds to the main branch of the NFRC at $F=0.5$N. (a) Amplitude and (b) phase lag.}
\label{fig_BBs_12}
\end{figure}

The identification of the 1:2 subharmonic resonance in Figure \ref{fig_BBs_12} is achieved in a similar manner. The transfer process from the rest position to the 1:2 isola is represented for $F=0.5$N in Figure \ref{fig_time_12}. In the present case, the system is attracted to a point relatively far away from the lower resonance point and then travels along the isola up to the upper resonance point. The backbone curve can be subsequently identified by setting $\Phi_\mathrm{ref}=-3\pi/8$ and decreasing the forcing amplitude $F$, as displayed for $F=0.8$N in Figure \ref{fig_BBs_12}.

\begin{figure}[htb!]
\begin{center}
\includegraphics[width=0.45\textwidth]{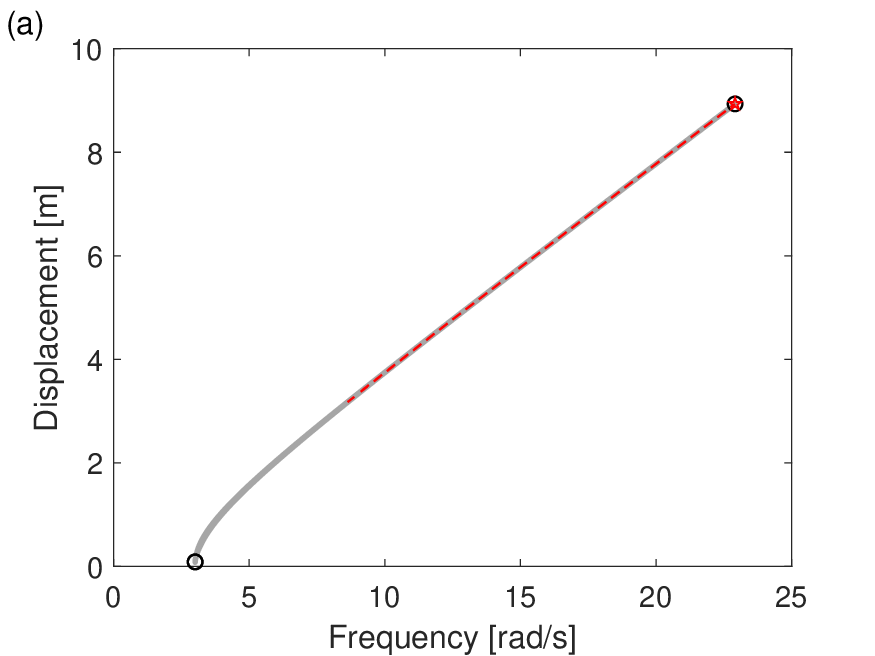} 
\includegraphics[width=0.45\textwidth]{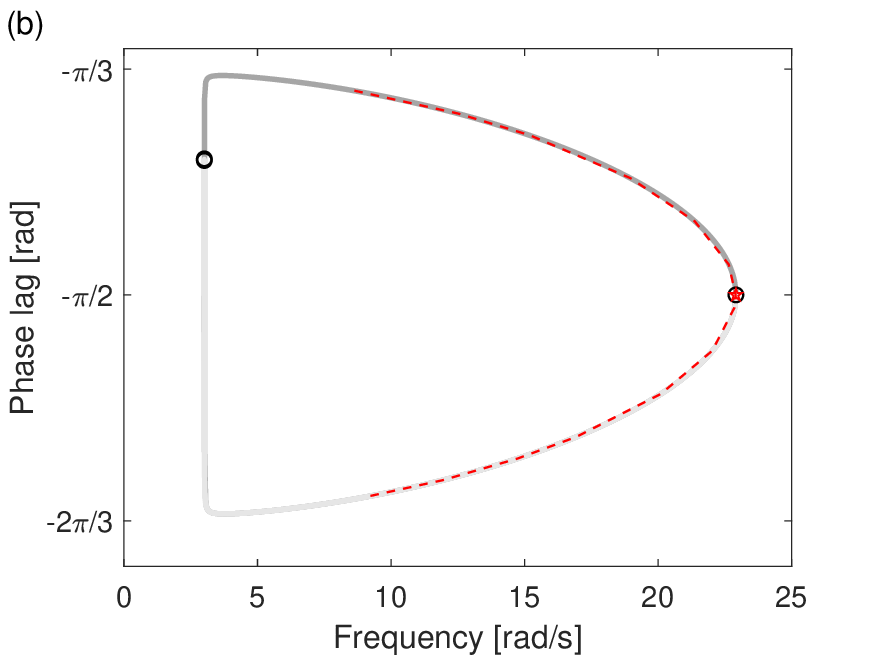}  
\end{center}
\caption{NFRC (\tpf{red}) of the 1:3 resonance identified by PLLT ($F=0.5$N). The starting point is the red marker. The reference solution is provided by HBM in gray; light gray corresponds to unstable responses. The bifurcation points are represented by circles. (a) Amplitude and (b) phase lag.}
\label{fig_FRCs_13}
\end{figure}

\begin{figure}[htb!]
\begin{center}
\includegraphics[width=0.45\textwidth]{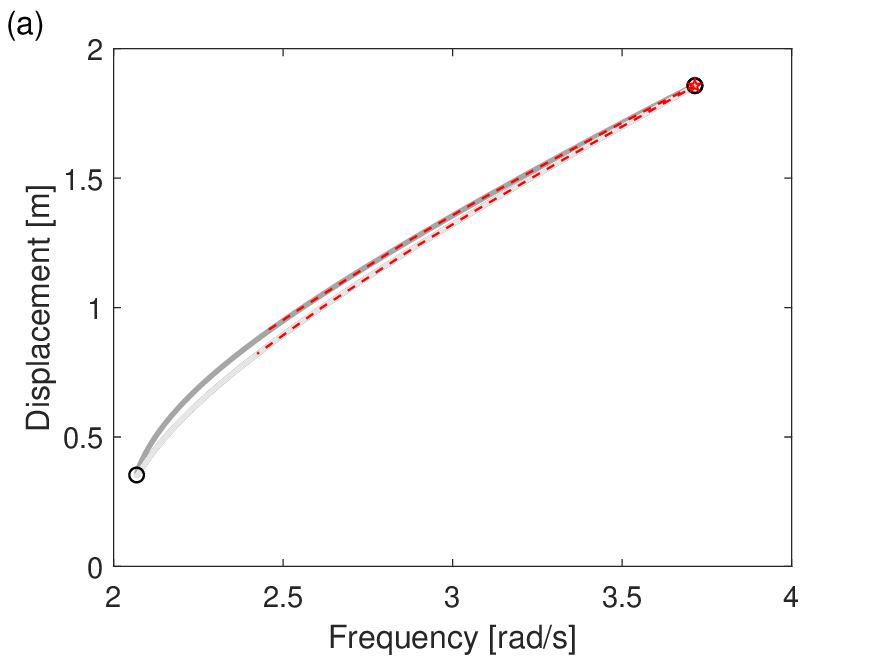} 
\includegraphics[width=0.45\textwidth]{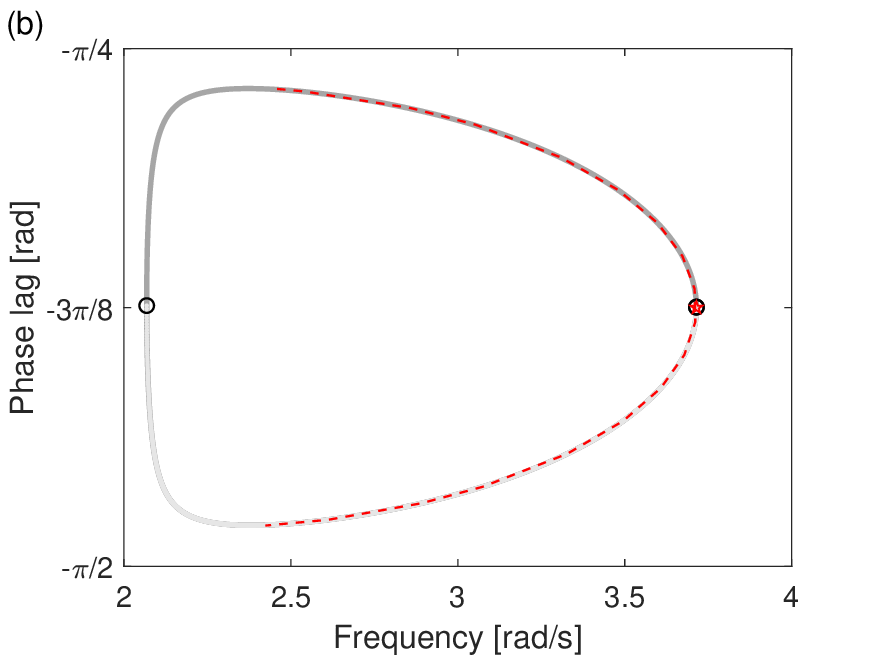}  
\end{center}
\caption{NFRC (\tpf{red}) of the 1:2 resonance identified by PLLT ($F=0.5$N). The starting point is the red marker. The reference solution is provided by HBM in gray; light gray corresponds to unstable responses. The bifurcation points are represented by circles. (a) Amplitude and (b) phase lag.}
\label{fig_FRCs_12}
\end{figure}

Large portions of the NFRC of the 1:3 and 1:2 subharmonic resonances can also be identified by modifying $\Phi_\mathrm{ref}$ from the upper resonance point while keeping $F$ constant. The upper and lower branches are obtained by either increasing or decreasing the reference phase lag, respectively. The results for $F=0.5$ N are presented in Figures \ref{fig_FRCs_13} and \ref{fig_FRCs_12}. The integral control gain has be increased to $K_I=0.05$ to enhance PLL stabilization and accelerate convergence in order to obtain a larger portion of the 1:3 resonance.

%==================================================================
\section{Experimental demonstration using a clamped-clamped beam}
%==================================================================
\subsection{Experimental setup}

\begin{figure}[htb!]
\begin{center}
\includegraphics[width=0.6\textwidth]{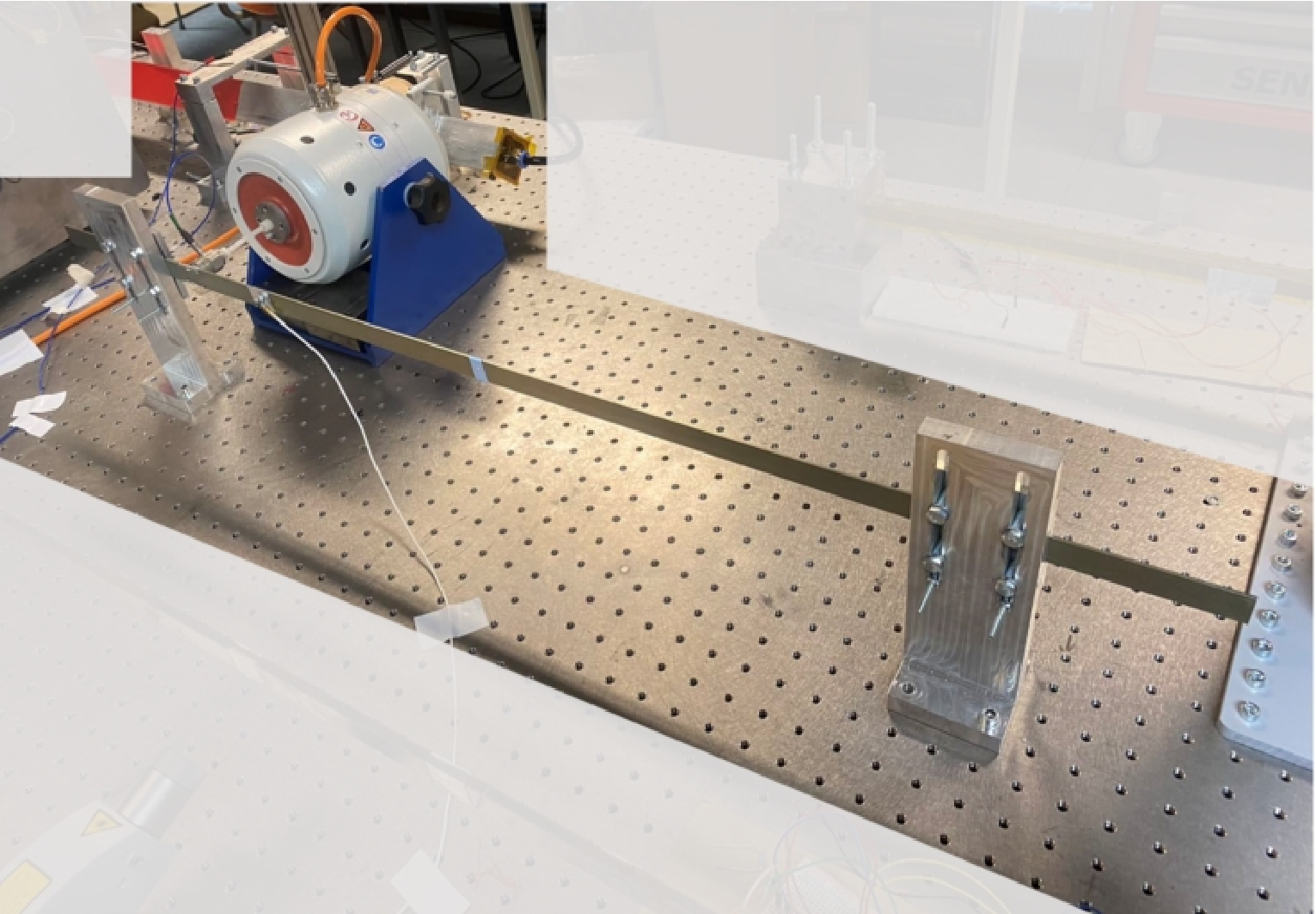}  
\end{center}
\caption{Experimental setup. }
\label{fig:exp_set_up1}
\end{figure}

The test sample is a thin straight aluminum beam of dimension $2\mathrm{mm}\times20\mathrm{mm}\times1\mathrm{m}$, see Figure~\ref{fig:exp_set_up1}. Both ends of the beam were clamped using two aluminum blocks bolted to a support frame fixed on a vibration table. The clamped beam length could be adjusted by changing the position of the aluminum blocks. A sliding groove of cross section $0.2\mathrm{mm}\times20\mathrm{mm}$ was made in the block to prescribe the beam's movements. 

During the installation process, low initial tension or compression force along the beam was ensured so as to avoid structural buckling. 
Such a slender beam is expected to exhibit geometrically nonlinear behavior resulting in an hardening effect due to mid-plane stretching. The first vibration mode is considered in this study. The formulas given for secondary resonances in \cite{nayfeh2008} were useful for guiding experiment design.

An electromagnetic shaker (TIRA TV 51075) fed by a power amplifier (TIRA BAA 120) working in current mode was used to drive the beam into vibration. The shaker was connected to the beam through a flexible stinger and an impedance head (DYTRAN 5860B). The shaker was placed close to one clamped boundary in order to minimize shaker–structure interaction. An accelerometer (DYTRAN 3035G) was placed closer to the beam center. The measured signals were amplified by a signal conditioner (PCB PIEZOTRONICS 482C). 

\begin{figure}[htb!]
\begin{center}
\includegraphics[width=0.7\textwidth]{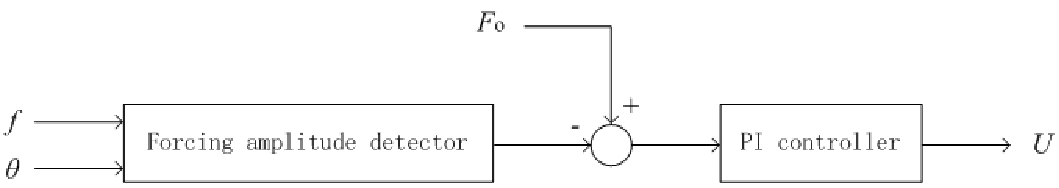}  
\end{center}
\caption{Fundamental force controller. }
\label{exp_set_up}
\end{figure}

PLLT was realized using the real-time controller dSPACE MicroLabBox with the control scheme implemented in Matlab/Simulink. The sampling frequency of the fixed-step solver was set to be 10 kHz. An additional controller was used to control the applied excitation toward a settled magnitude $F_o$ \cite{abeloos2022}, as schematized in Figure~\ref{exp_set_up}. The gains of the PI controller were both set to 0.01. The adaptive filter parameter $\mu$ was set to 0.001.

Despite its apparent simplicity, this set-up posed two important difficulties. First, shaker-structure interaction remained an important issue for PPLT with the consequence that it was not possible to find a single beam length for which all the resonances of interest could be identified robustly. Second, the set-up exhibited important day-to-day variability induced by the key role played by room temperature on the dynamical properties of the beam.

\subsection{Beam 1 (length=0.678m): primary and 1:2 subharmonic resonances}

\begin{figure}[htb!]
\begin{center}
\includegraphics[width=0.45\textwidth]{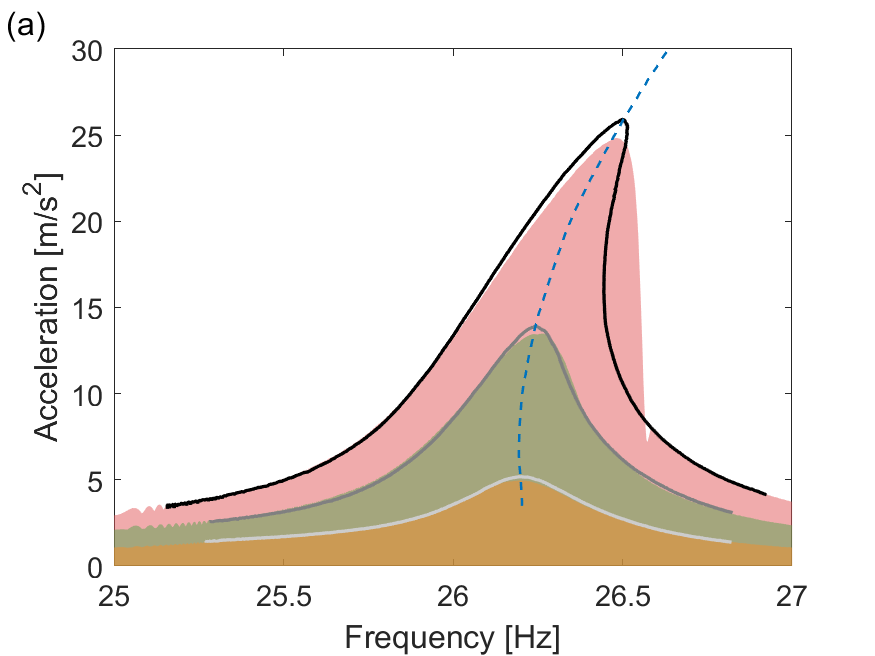} 
\includegraphics[width=0.45\textwidth]{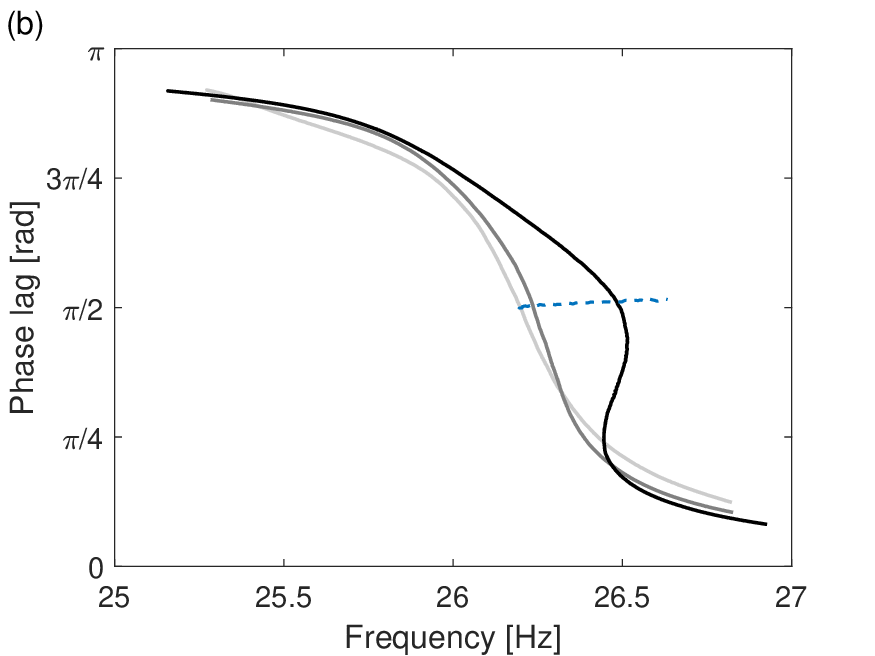}  
\end{center}
\caption{NFRCs (solid lines in grayscale) and  backbone (blue dashed line) of the primary resonance identified by PLLT using $K_P=1$ and $K_I=2$ at 0.2N, 0.4N and 0.6N. The colored areas correspond to SST. (a) Amplitude and (b) phase lag.}
\label{fig_exp_11}
\end{figure}

The NFRCs obtained from classical sine sweep testing (SST) and from PLLT around the primary resonance as well as the backbone identified by PLLT with $\Phi_\mathrm{ref}=\pi/2$ are displayed in Figure \ref{fig_exp_11}. Unlike the previous section, the phase lag was measured between the acceleration and forcing signals, so there is a difference of $\pi$ with respect to the results of Section 3. The beam is seen to exhibit slight softening behavior at low forcing levels followed by a more prominent hardening behavior at higher levels. The softening behavior can be attributed to the clampings whereas the hardening behavior is due to the geometrical nonlinearity. In addition to a good correspondence between the backbone and the local maxima in the NFRCs identified using PLLT,
there is also good agreement between the NFRCs obtained using PLLT and SST. The only difference is at $0.6$N where a jump is observed for SST due to the folded nature of the NFRC. Conversely, PLLT is able to unfold the NFRC and measure the complete branch including its unstable portion. We note that, in Figure \ref{fig_exp_11}(b), the measured phase lag is not exactly equal to $\pi/2$, which can be explained by the system's variability between different forcing levels as well as the existence of transients. However, the relative error between $\hat{\Phi}_{1,1}$ and $\Phi_\mathrm{ref}$ remained consistently below $2\%$.

\begin{figure}[htb!]
\begin{center}
\includegraphics[width=0.45\textwidth]{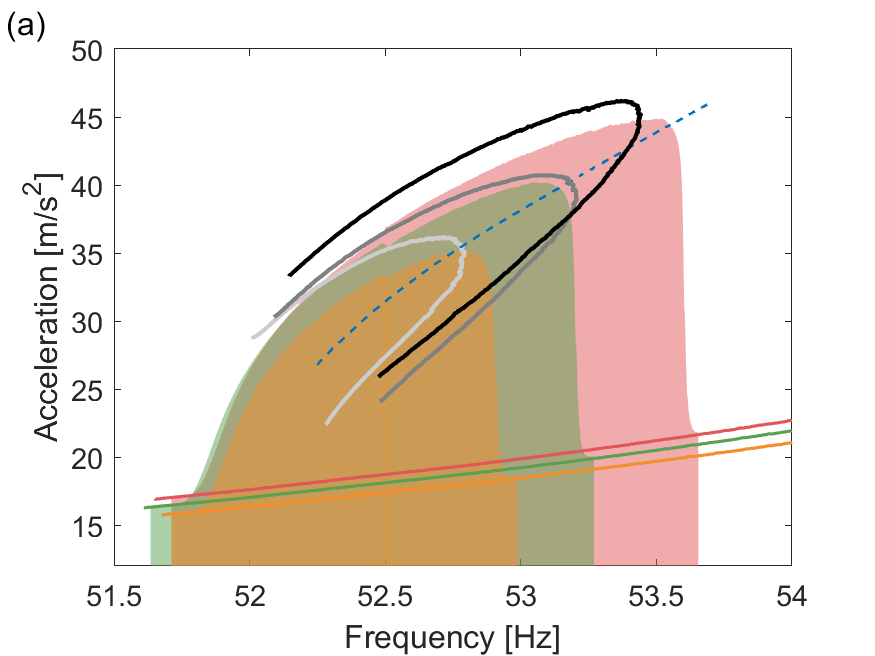} 
\includegraphics[width=0.45\textwidth]{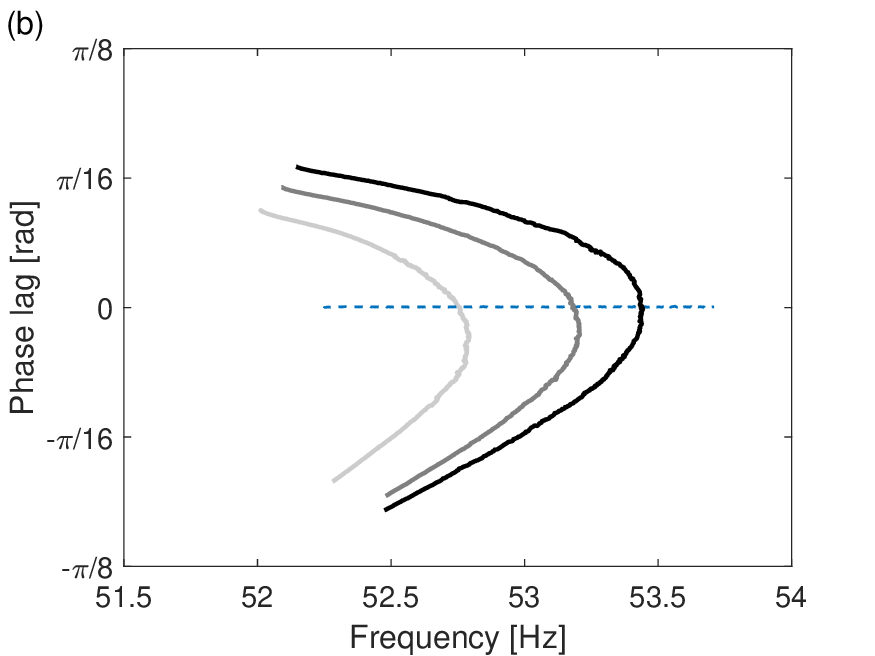}  
\end{center}
\caption{NFRCs (solid lines in grayscale) and  backbone (blue dashed line) of the 1:2 subharmonic resonance identified by PLLT using $K_P=10$ and $K_I=25$ at 2.5N, 2.6N and 2.7N. The colored areas correspond to SST. (a) Amplitude and (b) phase lag.}
\label{fig_exp_12}
\end{figure}

\begin{figure}[htb!]
\begin{center}
\includegraphics[width=0.45\textwidth]{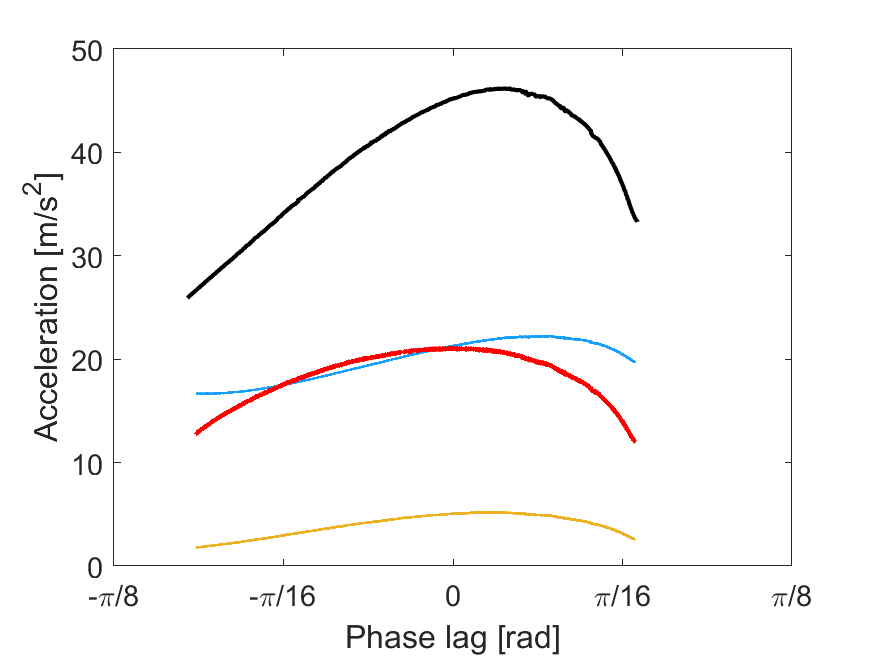} 
\end{center}
\caption{Different harmonics of the 1:2 subharmonic resonance measured at 2.7N: $\kappa=1$ (red), $\kappa=2$ (blue) and $\kappa=3$ (yellow). The total amplitude is represented in black.}
\label{fig_exp_Harm_12}
\end{figure}

To verify the existence of the 1:2 subharmonic isola, an impact hammer was used to trigger the jump from the main branch to the isola while the shaker was exciting the structure at approximately twice the primary resonance frequency. The isola was found to exist for $F \geq 2.3 $N. The starting point on the isola for forcing amplitudes between 2.5N and 2.7N was also obtained using the impact hammer. The NFRCs measured with PLLT are depicted in Figure \ref{fig_exp_12}; the different harmonics measured at 2.7N are plotted in Figure \ref{fig_exp_Harm_12}. Large PLL controller gains, i.e., $K_P=10$ and $K_I=25$, were necessary during the identification process. From the starting point, the NFRCs were obtained by sequentially decreasing the phase lag, resulting in an increase then decrease in forcing frequency. This strategy allowed us to trace the high-amplitude portion of the 1:2 isola around the turning point. 

The backbone in Figure \ref{fig_exp_12} was identified using $\Phi_\mathrm{ref}=0$, which is not in accordance with the theoretical expectations for a Duffing-like type of behavior. A possible explanation for this discrepancy is that the beam has a non-negligible initial curvature resulting in a Helmholtz-like behavior for which the resonant phase lag between the acceleration and forcing signals of a 1:2 subharmonic resonance is 0 or $\pi$ \cite{volvertthesis}. Nonetheless, the backbone is found to provide a good approximation to the locus of the local maxima on the 1:2 isolas. Finally, the results of SST are also given in Figure \ref{fig_exp_12}. A transition to the isola during SST was possible by applying perturbations. The agreement between PLLT and SST is very good considering the challenges posed by the inherent variability of the set-up and shaker-structure interactions.

\subsection{Beam 2 (length=0.754m): 2:1/3:1 superharmonic and 1:2 subharmonic resonances}

\begin{figure}[htb!]
\begin{center}
\includegraphics[width=0.45\textwidth]{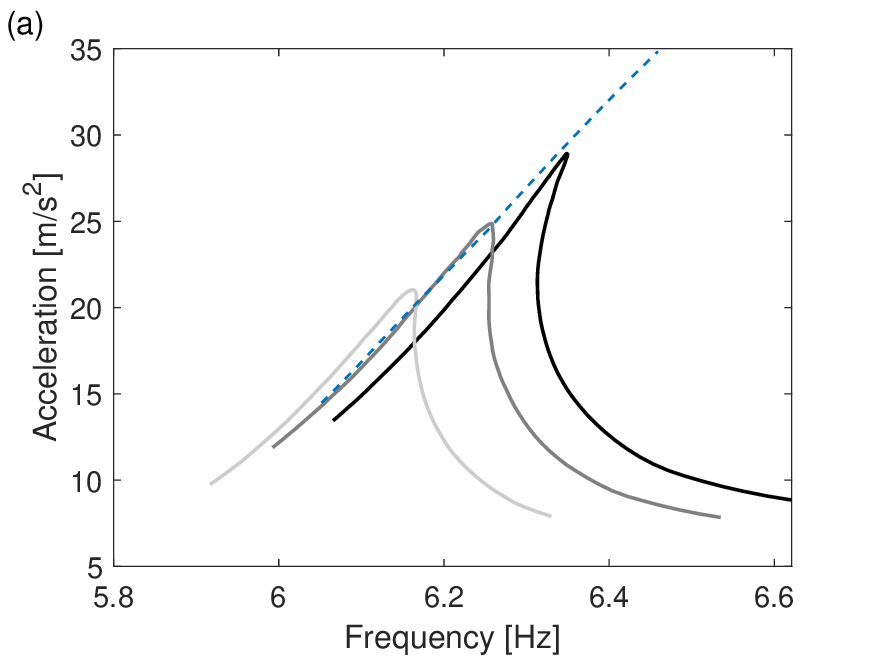} 
\includegraphics[width=0.45\textwidth]{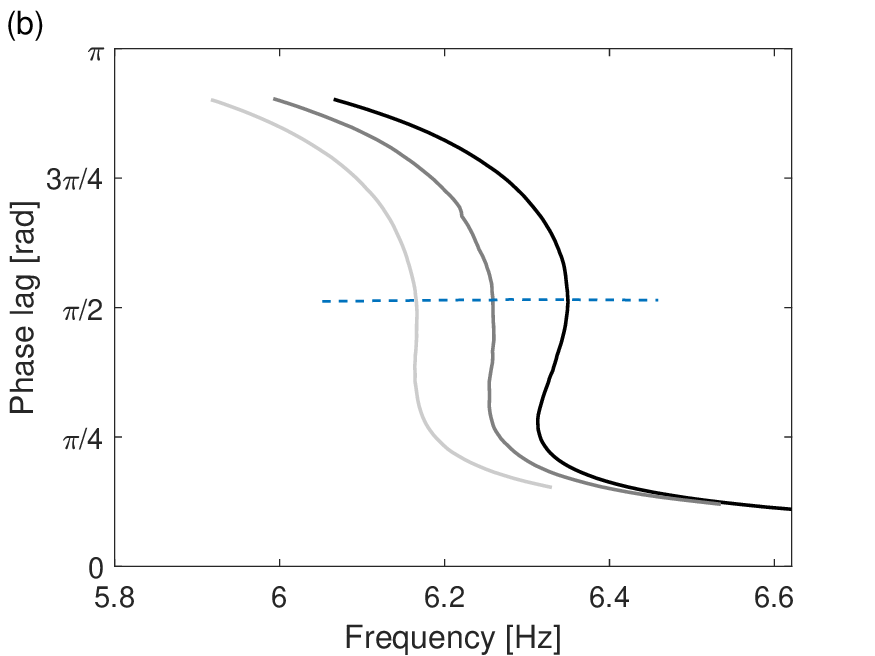}  
\end{center}
\caption{NFRCs (solid lines in grayscale) and backbone (blue dashed line) of the 3:1 superharmonic resonance identified by PLLT using 
$K_P=0.1$ and $K_I=0.05$ at 3.5N, 4.0N and 4.5N. (a) Amplitude and (b) phase lag.}
\label{fig_exp_31}
\end{figure}

\begin{figure}[htb!]
\begin{center}
\includegraphics[width=0.45\textwidth]{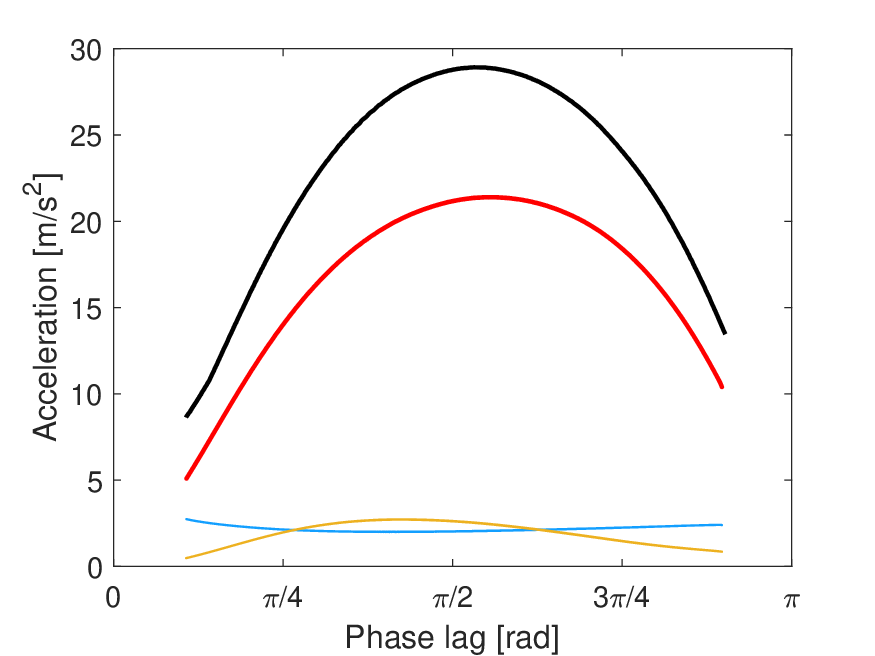}  
\end{center}
\caption{Different harmonics of the 3:1 superharmonic resonance measured at 4.5N: $\kappa=1$ (blue), $\kappa=3$ (red) and $\kappa=5$ (yellow). The total amplitude is represented in black.}
\label{fig_exp_Harm_31}
\end{figure}

The results of the identification of the 3:1 superharmonic resonance are shown in Figure~\ref{fig_exp_31}. The backbone obtained by locking the phase at the theoretically-expected value of $\pi/2$ for the acceleration ($-\pi/2$ for the displacement) provides an excellent approximation of the local maxima of the three NFRCs. The different harmonics measured at 4.5N in Figure \ref{fig_exp_Harm_31} indicates that the third harmonic in the vicinity of the frequency of the primary resonance is the dominant harmonic.

%3rd harmonic amplitude and phase angle

%3:1 relation in time signals
%time domain comparison and small force distortion
%time trace shows force distortion

\begin{figure}[htb!]
\begin{center}
\includegraphics[width=0.45\textwidth]{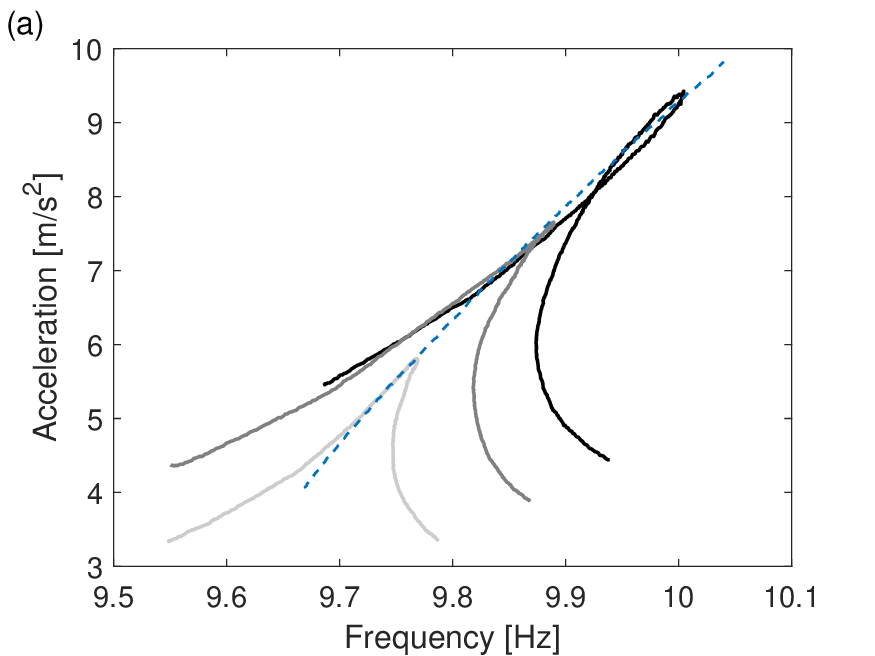} 
\includegraphics[width=0.45\textwidth]{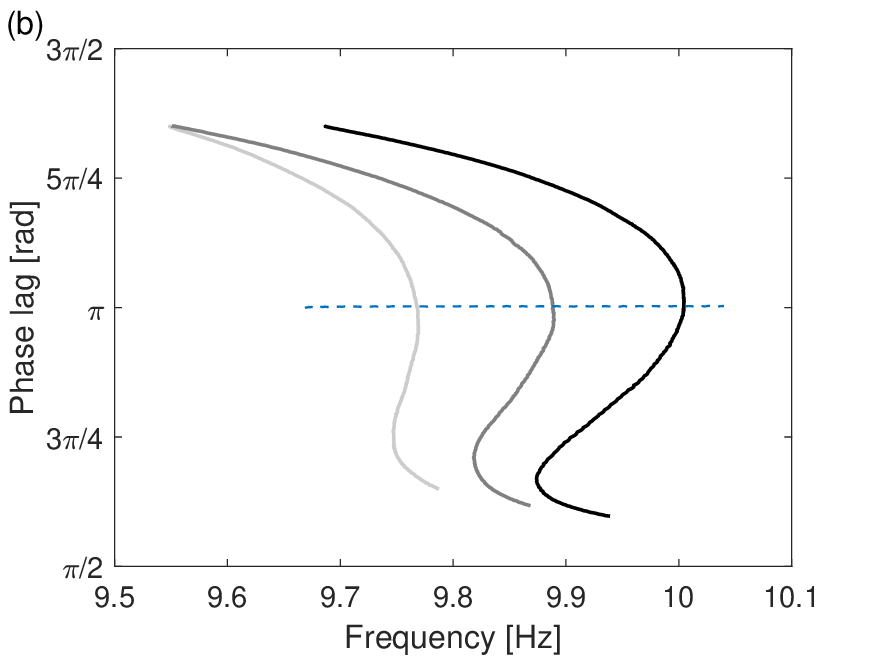} 
\end{center}
\caption{NFRCs (solid lines in grayscale) and backbone (blue dashed line) of the 2:1 superharmonic resonance identified by PLLT using 
$K_P=1$ and $K_I=2$ at 2.5N, 3.0N and 3.5N. (a) Amplitude and (b) phase lag.}
\label{fig_exp_21_2}
\end{figure}

The identification of 2:1 superharmonic resonance was performed in a similar manner. Despite that a variability of the resonance frequency was observed during PLLT, a good correspondence between the NFRCs and the backbone curve was obtained in Figure \ref{fig_exp_21_2}.

%Amplitude (phase) resonances at different force levels occur around the same phase lag value (0.24).
%The folding bifurcations take places when the phase lags are around 0.
%The deviation with the theoretical values are possibly stems from the quadratic stiffness term induced by beam curvature.
%It is generated due to misalignment of boundary bases, resulting in curvature in the long test beam.
%This is embodied by time domain responses of primary resonance testing and the dependence of the linear resonance frequency of the first mode on base installation angles (perpendicular to vibration table).

%\begin{figure}[htb!]
%\begin{center}
%\includegraphics[width=0.45\textwidth]{figures/z_21_1_pll_frc.eps} 
%\end{center}
%\caption{NFRCs (solid lines in grayscale) and backbone (dashed lines) of the 2:1 superharmonic resonance identified by PLLT using 
%$K_P=1$ and $K_I=2$ at 3.0N, 3.5N and 4.0N.}
%\label{fig_exp_2113}
%\end{figure}

\begin{figure}[htb!]
\begin{center}
\includegraphics[width=0.45\textwidth]{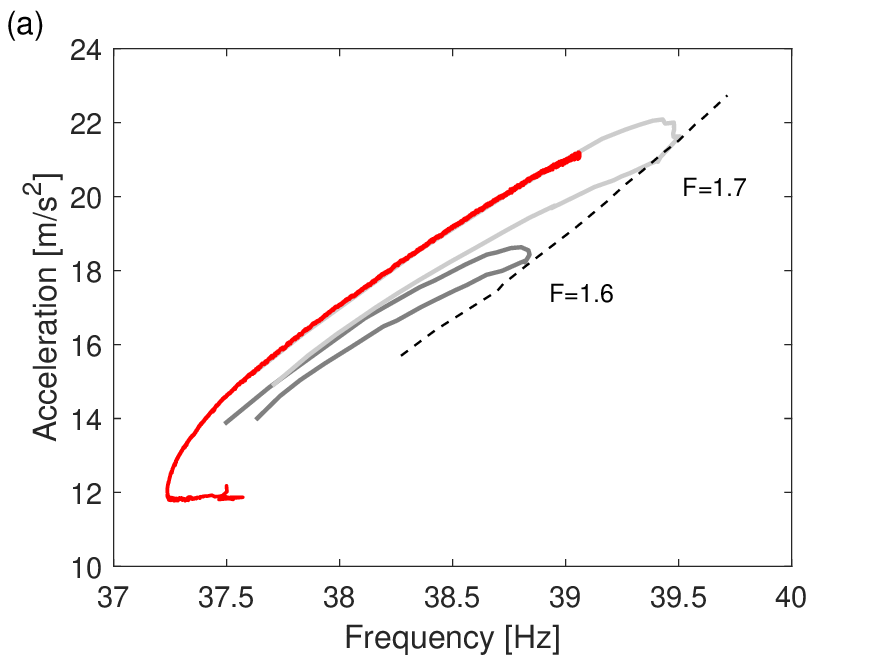} 
\includegraphics[width=0.45\textwidth]{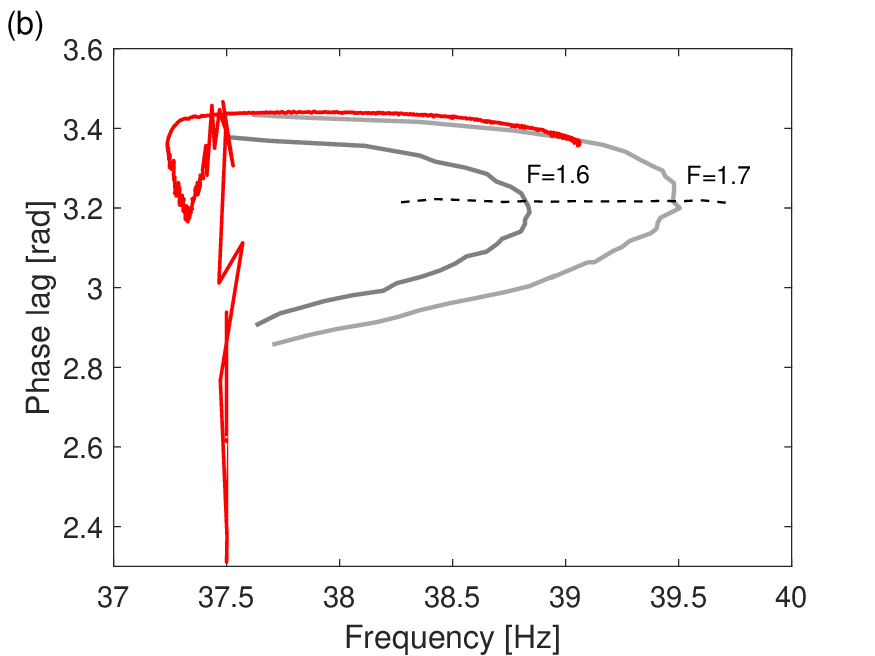}  
\includegraphics[width=0.45\textwidth]{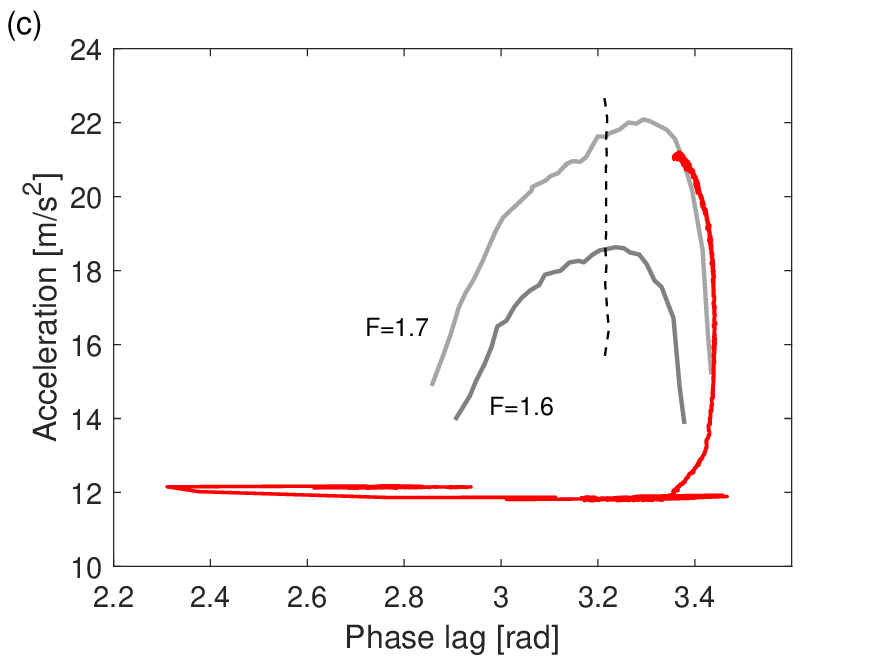} 
\includegraphics[width=0.45\textwidth]{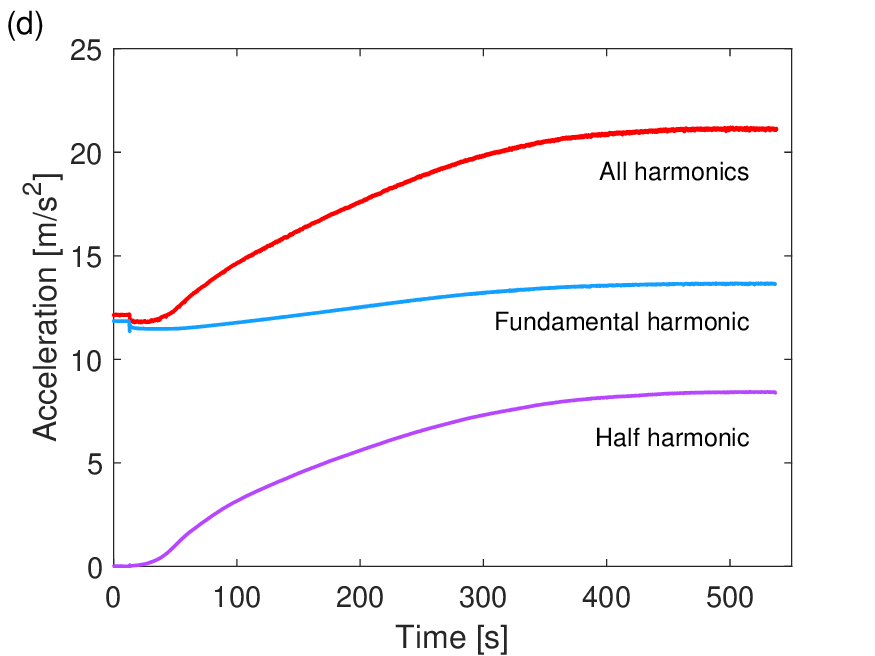}  
\end{center}
\caption{Automatic transition to the detached 1:2 subharmonic resonance and subsequent identification of the NFRCs and backbone. State transfer: $K_P=1$ and $K_P=0.5$; identification: $K_P=5$ and $K_P=20$.}
\label{fig_exp_climb}
\end{figure}

Finally, the automatic transfer to the 1:2 isola and the subsequent identification were also successfully achieved for beam 2. As shown in Figure~\ref{fig_exp_climb}, after some transient oscillations, the system was successfully attracted by the 1:2 resonance and travelled along the isolated branch. Figure~\ref{fig_exp_climb}(d) illustrates the growth of the half harmonic during the transfer from the main branch to the subharmonic orbit.

%==================================================================
 \section{Conclusions}
%==================================================================

In this study, a control-based vibration testing scheme for identifying both primary and secondary resonances of nonlinear systems was proposed. The PLLT scheme integrates digital adaptive filters into the phase detector module of PLL. The experimental continuation of NFRCs exploited the fact that, unlike the forcing frequency, the phase lag has a monotonic evolution around nonlinear resonances. Thus, the NFRC can be obtained by sweeping the phase lag at constant forcing amplitude. The experimental continuation of backbone curves exploited the concept of a resonant phase lag. In this case, the forcing amplitude is swept at constant phase lag. 

The performance of PLLT was carefully assessed using numerical and physical experiments. The NFRCs (with their stable and unstable portions) and backbones of different superharmonic and subharmonic resonances were successfully identified. A salient feature of PLLT is that both bifurcating and isolated branches were reached from the rest position without any user interaction. Throughout this study, the selection of the gains of the PI controller remained a matter of trial and error. Further research concerning this aspect is necessary.

%==================================================================

%==================================================================
	
	\bibliographystyle{unsrt}
	\bibliography{main}

\end{document}